\documentclass[preprint,preprintnumbers,amsmath,amssymb]{revtex4}

\usepackage{bm}
\pdfoutput=1
\usepackage{graphicx}
\usepackage{subfig}
\usepackage{array}
\usepackage{color}
\usepackage{ulem}

\graphicspath{{figures/}}

\makeatletter

\newcommand{\Rmnum}[1]{\expandafter\@slowromancap\romannumeral #1@}
\makeatother

\begin{document}

\title{On the four-zero texture of quark mass matrices and its stability}

\author{Zhi-zhong Xing$^{a,b}$}\email{xingzz@ihep.ac.cn}\quad
\author{Zhen-hua Zhao$^a$}\email{zhaozhenhua@ihep.ac.cn}

\affiliation{$^a$Institute of High Energy Physics, Chinese Academy
of Sciences, Beijing 100049, China \\
$^b$Center for High Energy Physics, Peking University, Beijing
100080, China}

\begin{abstract}
We carry out a new study of quark mass
matrices $M^{}_{\rm u}$ (up-type) and $M^{}_{\rm d}$ (down-type)
which are Hermitian and have four zero entries, and find a new part
of the parameter space which was missed in the previous works. We
identify two more specific four-zero patterns of $M^{}_{\rm u}$
and $M^{}_{\rm d}$ with fewer free parameters, and present two toy
flavor-symmetry models which can help realize such special and
interesting quark flavor structures. We also show that the
texture zeros of $M^{}_{\rm u}$ and $M^{}_{\rm d}$ are essentially
stable against the evolution of energy scales in an analytical
way by using the one-loop renormalization-group equations.
\end{abstract}

\maketitle

\section{Introduction}

The discovery of the Higgs boson \cite{H} signifies the ``completion" of the
Standard Model (SM) which is not only phenomenologically successful
but also theoretically self-consistent. In particular, it verifies
the Brout-Englert-Higgs mechanism and Yukawa interactions which are
responsible for the generation of lepton and quark masses. However,
the SM is not really ``complete" in the sense that it cannot explain
the origin of neutrino masses, the structures of lepton and quark
flavors, the asymmetry of matter and antimatter in the Universe, the
nature of dark matter, etc. Hence one has to go beyond the SM and
explore possible new physics behind it in order to solve the
aforementioned puzzles.

Here let us focus on the flavor puzzles in the SM. The flavor issues
mainly refer to the generation of fermion masses, the dynamics of
flavor mixing and the origin of CP violation. Even within the SM in
which all the neutrinos are assumed to be massless, there are
thirteen free flavor parameters which have to be experimentally
determined. On the other hand, one is also puzzled by the observed
spectra of lepton and quark masses and the observed patterns of
flavor mixing, which must imply a kind of underlying flavor
structure \cite{Xing2014}.

In this paper we restrict ourselves to the flavor issues in the
quark sector where there are ten free parameters: six quark masses,
three flavor mixing angles and one CP-violating phase. Thanks to the
coexistence of Yukawa interactions and charged-current gauge
interactions, the flavor and mass bases of three quark families do
not coincide with each other, leading to the phenomenon of flavor
mixing and CP violation. The latter is described by a $3\times 3$
unitary matrix $V$, the so-called Cabibbo-Kobayashi-Maskawa (CKM)
matrix \cite{CKM},
\begin{eqnarray}\label{CKM}
V = \left( \begin{matrix} V^{}_{ud} & V^{}_{us} & V^{}_{ub} \cr
V^{}_{cd} & V^{}_{cs} & V^{}_{cb} \cr V^{}_{td} & V^{}_{ts} &
V^{}_{tb} \cr \end{matrix} \right) \; ,
\end{eqnarray}
which can be parameterized in terms of three mixing angles
($\theta^{}_{12}, \theta^{}_{13}, \theta^{}_{23}$) and one
CP-violating phase ($\delta$) via the definitions of $V^{}_{us}
= \sin\theta^{}_{12} \cos\theta^{}_{13}$, $V^{}_{ub} =
\sin\theta^{}_{13} e^{-{\rm i} \delta}$ and $V^{}_{cb} =
\cos\theta^{}_{13} \sin\theta^{}_{23}$ and the unitarity of $V$
itself. As $V$ originates from a mismatch between the
diagonalizations of the up-type quark mass matrix $M^{}_{\rm u}$ and
the down-type one $M^{}_{\rm d}$, which are equivalent to
transforming their flavor bases into their mass bases, an attempt to
calculate the flavor mixing parameters should start from the mass
matrices in the flavor basis. In view of the experimental results
$m^{}_{u}\ll m^{}_{c}\ll m^{}_{t}$, $m^{}_{d}\ll m^{}_{s}\ll
m^{}_{b}$ and $\theta^{}_{13}\ll \theta^{}_{23} \ll \theta^{}_{12}
\equiv \theta^{}_{\rm C} \simeq 13^\circ$, where $\theta^{}_{\rm C}$
denotes the Cabibbo angle, we believe that the strong hierarchy of
three flavor mixing angles must be attributed to the strong
hierarchy of quark masses.

Therefore, one is tempted to relate the smallness of three flavor
mixing angles with the smallness of four independent mass ratios
$m^{}_u/m^{}_c$, $m^{}_c/m^{}_t$, $m^{}_d/m^{}_s$ and
$m^{}_s/m^{}_b$. A famous relation of this kind is the
Gatto-Sartori-Tonin (GST) relation $\sin\theta^{}_{12}\sim
\sqrt{m^{}_d/m^{}_s}$ \cite{gst}. The Fritzsch ansatz of quark mass
matrices \cite{fritzsch},
\begin{equation}\label{fritzsch}
 M^{\rm F}_{{\rm u}}=\left(
 \begin{array}{ccc}
 0& C^{}_{{\rm u}} &0\\
C_{{\rm u}}^* & 0 &  B^{}_{{\rm u}}\\
0& B_{{\rm u}}^* & A^{}_{{\rm u}}
 \end{array}
 \right) \; ,
 \hspace{1.5cm}
  M^{\rm F}_{{\rm d}}=\left(
 \begin{array}{ccc}
 0& C^{}_{{\rm d}} &0\\
C_{{\rm d}}^* & 0 &  B^{}_{{\rm d}}\\
0& B_{{\rm d}}^* & A^{}_{{\rm d}}
 \end{array}
 \right) \; ,
\end{equation}
can easily lead us to the above GST relation. Note that $M^{\rm
F}_{{\rm u}}$ and $M^{\rm F}_{{\rm d}}$ possess the parallel structures
with the same zero entries. Furthermore, they have been taken to be
Hermitian without loss of generality, since a rotation of the
right-handed quark fields does not affect any physical results in
the SM or its extensions which have no flavor-changing right-handed
currents. The Fritzsch ansatz totally involves eight independent
parameters, and thus it can predict two relations among six quark
masses and four flavor mixing parameters. However, it has been shown
that this simple ansatz is in conflict with current experimental
data \cite{sixzero}.

One may modify the Fritzsch ansatz by reducing the number of its
texture zeros. Given a Hermitian or symmetric mass matrix, a pair of
its off-diagonal texture zeros are always counted as one zero. Hence
the Fritzsch ansatz has six nontrivial texture zeros. It has been shown
that adding nonzero (1,1) or (1,3) entries to $M^{\rm F}_{\rm u}$
and $M^{\rm F}_{\rm d}$ does not help much \cite{Shrock}, but the
following Fritzsch-like ansatz is phenomenologically viable
\cite{fourzero,fourzero2}:
\begin{equation}\label{four}
 M^{}_{{\rm u}}=\left(
 \begin{array}{ccc}
 0& C^{}_{{\rm u}} &0\\
C_{{\rm u}}^* & \tilde B^{}_{{\rm u}} &  B^{}_{{\rm u}}\\
0& B_{{\rm u}}^* & A^{}_{{\rm u}}
 \end{array}
 \right) \; ,
 \hspace{1.5cm}
  M^{}_{{\rm d}}=\left(
 \begin{array}{ccc}
 0& C^{}_{{\rm d}} &0\\
C_{{\rm d}}^* & \tilde B^{}_{{\rm d}} &  B^{}_{{\rm d}}\\
0& B_{{\rm d}}^* & A^{}_{{\rm d}}
 \end{array}
 \right) \; .
\end{equation}
We see that Hermitian $M^{}_{{\rm u}}$ and $M^{}_{{\rm d}}$ have the
up-down parallelism and four texture zeros. So far a lot of interest
has been paid to the phenomenological consequences of Eq. (3)
\cite{fourzero,fourzero2,previous,recent}. In particular, the
parameter space of this ansatz was numerically explored in Ref.
\cite{xingzhang}, where a mild hierarchy $|B|/A\sim \tilde B/|B|
\sim 0.24$ was found to be favored for both up and down sectors.
Here we have omitted the subscript ``\rm u" and ``\rm d" for the
relevant parameters, and we shall do so again when discussing
something common to $M^{}_{{\rm u}}$ and $M^{}_{{\rm d}}$ throughout
this paper. Although there are four complex parameters in Eq. (3),
only two linear combinations of the four phases are physical and can
simply be denoted as $\phi^{}_1 = \arg(C^{}_{\rm u}) -
\arg(C^{}_{\rm d})$ and $\phi^{}_2 = \arg(B^{}_{\rm u}) -
\arg(B^{}_{\rm d})$. It has been found that $\phi^{}_2$ is very
close to zero or $2\pi$ \cite{xingzhang}, while $\sin{\phi^{}_1}$ is
close to $\pm 1$ and its sign can be fixed by $\eta^{}_{\rm
u}\eta^{}_{\rm d}\sin{\phi^{}_1}>0$, where the dimensionless
coefficients $\eta^{}_{\rm u}$ and $\eta^{}_{\rm d}$ will be defined
in section II.

In this paper we aim to carry out a new study of the four-zero
texture of quark mass matrices and improve the previous works in
the following aspects:
\begin{itemize}
\item     We reexplore the parameter space of $M^{}_{\rm u}$
and $M^{}_{\rm d}$ by taking into account the updated values of
quark masses and the latest results of the CKM flavor mixing
parameters. The new analysis leads us to a new part of the parameter
space, which is interesting but was missed in Ref. \cite{xingzhang}
and other references.

\item     We identify two more
specific four-zero patterns of $M^{}_{\rm u}$ and $M^{}_{\rm d}$
with fewer free parameters. Namely, there is a kind of parameter
correlation in such an ansatz, making the exercise of model building
much easier. We present two toy flavor-symmetry models to realize
such special and interesting quark flavor structures.

\item     The running behaviors of $M^{}_{\rm u}$ and $M^{}_{\rm d}$
from a superhigh scale down to the electroweak scale are
studied in an analytical way by using the one-loop
renormalization-group equations (RGEs), in order to examine whether
those texture zeros are stable against the evolution of energy
scales. We find that they are essentially stable in the SM.
\end{itemize}
The remaining parts of this paper are organized as follows. In
section II we first explore the complete parameter space of $M^{}_{\rm u}$
and $M^{}_{\rm d}$ and then discuss the relevant phenomenological
consequences. Particular attention will be paid to some properties of
the four-zero texture that the previous works did not put emphasis on.
Section III is devoted to discussions about the special patterns of
four-zero quark mass matrices in which some particular relations
among the finite matrix elements are possible. Two toy
flavor-symmetry models, which can help realize such interesting
patterns, will be presented for the sake of illustration. In section
IV we derive the one-loop RGE corrections to
$M^{}_{\rm u}$ and $M^{}_{\rm d}$ which evolve from a superhigh
energy scale down to the electroweak scale. Our analytical results
show that those texture zeros are essentially stable against the
evolution of energy scales. As a byproduct, the possibility of applying
the four-zero texture of quark mass matrices to resolving the strong
CP problem is also discussed in a brief way. Finally, we summarize
our main results and make some concluding remarks on the quark
flavor issues in section V.

\section{The parameter space: results and explanations}

Before performing an updated and complete numerical analysis of the
parameter space of Hermitian $M^{}_{\rm u}$ and $M^{}_{\rm d}$ with
four texture zeros, let us briefly reformulate the relations between
the parameters of $M^{}_{\rm u, d}$ and the observable quantities
\cite{xingzhang}. First of all, $M^{}_{\rm u,d}$ can be transformed
into a real symmetric matrix $\overline M^{}_{\rm u,d}$ through a
phase redefinition:
\begin{equation}\label{}
\overline M = P^\dagger_{} M_{} P_{}=\left(
 \begin{array}{ccc}
 0&\hspace{0.3cm} |C_{{}}| &\hspace{0.3cm}0\\
|C_{{}}| &\hspace{0.3cm}\tilde B & \hspace{0.3cm} |B_{{}}|\\
0&\hspace{0.3cm} |B_{{}}| &\hspace{0.3cm} A_{{}}
 \end{array}
 \right) \; ,
\end{equation}
where the subscript ``u" or ``d" has been omitted, and $P={\rm
Diag}\{1, e^{-{\rm i} \phi^{}_{C}} , e^{-{\rm
i}(\phi^{}_{C}+\phi^{}_{B})}\}$ with $\phi^{}_B = \arg(B)$ and
$\phi^{}_C = \arg(C)$. Of course, one may diagonalize $\overline M$
as follows:
\begin{equation}\label{masses}
O^{\rm T}\overline M_{}O = \left(
 \begin{array}{ccc}
 \lambda^{}_1&\hspace{0.3cm}  &\hspace{0.3cm}\\
 &\hspace{0.3cm}\lambda^{}_2 & \hspace{0.3cm}\\
&\hspace{0.3cm} &\hspace{0.3cm} \lambda^{}_3
 \end{array}
 \right) \; .
\end{equation}
Without loss of generality, we require $A$ and $\lambda^{}_3$ to be
positive. Then $|B|$, $\tilde{B}$ and $|C|$ can be expressed in
terms of $A$ and the three quark mass eigenvalues $\lambda^{}_i$
(for $i =1,2,3$, corresponding to $m^{}_u, m^{}_c, m^{}_t$ in the up
sector or $m^{}_d, m^{}_s, m^{}_b$ in the down sector):
\begin{equation}\label{parameter}
  \begin{array}{cll}
  \vspace{0.2cm}
  \tilde B &=& \lambda^{}_1+\lambda^{}_2+\lambda^{}_3-A\ ,\\
  \vspace{0.2cm}
  |B|&=&\sqrt{\displaystyle
  \frac{\left(A-\lambda^{}_1\right)\left(A-\lambda^{}_2\right)
  \left(\lambda^{}_3-A\right)}{A}}\ ,\\
  |C|&=&\sqrt{\displaystyle \frac{-\lambda^{}_1\lambda^{}_2\lambda^{}_3}{A}} \
  .
  \end{array}
\end{equation}
In this case the orthogonal matrix $O$ reads
\footnotesize
\begin{equation}\label{}
O = \left(
 \begin{array}{ccc}
 \vspace{0.3cm}
 \sqrt{\displaystyle\frac{\lambda^{}_2 \lambda^{}_3 \left(A-
 \lambda^{}_1\right)}
 {A\left(\lambda^{}_2-\lambda^{}_1\right)
 \left(\lambda^{}_3-\lambda^{}_1\right)}}&\hspace{0.5cm}
 \displaystyle\eta\sqrt{ \frac{\lambda^{}_1 \lambda^{}_3
 \left(\lambda^{}_2-A\right)}
 {A\left(\lambda^{}_2-\lambda^{}_1\right)
 \left(\lambda^{}_3-\lambda^{}_2\right)}}&\hspace{0.5cm}
 \displaystyle\sqrt{ \frac{\lambda^{}_1 \lambda^{}_2 \left(A-
 \lambda^{}_3\right)}
 {A\left(\lambda^{}_3-\lambda^{}_1\right)
 \left(\lambda^{}_3-\lambda^{}_2\right)}} \\
 \vspace{0.3cm}
\displaystyle -\eta\sqrt{\frac{\lambda^{}_1 \left(
\lambda^{}_1-A\right)}{\left(\lambda^{}_2-\lambda^{}_1\right)
 \left(\lambda^{}_3-\lambda^{}_1\right)}}&\hspace{0.5cm}
\displaystyle \sqrt{ \frac{\lambda^{}_2
\left(A-\lambda^{}_2\right)}{\left(\lambda^{}_2-\lambda^{}_1\right)
 \left(\lambda^{}_3-\lambda^{}_2\right)}}&\hspace{0.5cm}
\displaystyle \sqrt{ \frac{\lambda^{}_3 \left(
\lambda^{}_3-A\right)}{\left(\lambda^{}_3-\lambda^{}_1\right)
 \left(\lambda^{}_3-\lambda^{}_2\right)}} \\
  \vspace{0.3cm}
\displaystyle \eta
\sqrt{\frac{\lambda^{}_1\left(A-\lambda^{}_2\right)\left(A-\lambda^{}_3\right)}{A
 \left(\lambda^{}_2-\lambda^{}_1\right)\left(\lambda^{}_3-\lambda^{}_1\right)}}&
\displaystyle  -
\sqrt{\frac{\lambda^{}_2\left(A-\lambda^{}_1\right)\left(\lambda^{}_3-A\right)}{A
  \left(\lambda^{}_2-\lambda^{}_1\right)\left(\lambda^{}_3-\lambda^{}_2\right)}}&
\displaystyle
\sqrt{\frac{\lambda^{}_3\left(A-\lambda^{}_1\right)\left(A-\lambda^{}_2\right)}{A
\displaystyle\left(\lambda^{}_3-\lambda^{}_1\right)
\left(\lambda^{}_3-\lambda^{}_2\right)}}
 \end{array}
 \right) \; ,
\end{equation}
\normalsize where $\eta = \pm 1$, and the emergence of this
coefficient can be understood as follows. Since $A$ and
$\lambda^{}_3$ have been taken to be positive, $\lambda^{}_1$ and
$\lambda^{}_2$ must have the opposite signs so as to assure a
negative value of the determinant of $\overline M$,
\begin{equation}\label{}
  {\rm Det}(\overline M) = -A|C|^2=\lambda^{}_1\lambda^{}_2\lambda^{}_3 \
  .
\end{equation}
When identifying $\lambda^{}_{1,2,3}$ with the physical quark
masses, we use $\eta = +1$ and $-1$ to label the cases
$(\lambda^{}_1, \lambda^{}_2, \lambda^{}_3) = (-m^{}_u, m^{}_c,
m^{}_t)$ and $(\lambda^{}_1, \lambda^{}_2, \lambda^{}_3) = (m^{}_u ,
-m^{}_c, m^{}_t)$ in the up sector, respectively. The same labeling
is valid for the down sector.

In terms of quark mass eigenstates, the weak charged-current
interactions are written as
\begin{eqnarray}
-{\cal L}^{}_{\rm cc} = \frac{g_{2}^{}}{\sqrt{2}} ~ \overline{\left(u ~~ c
~~ t\right)^{}_{\rm L}} ~ \gamma^\mu \ V \left(\begin{matrix} d \cr
s \cr b \cr\end{matrix}\right)_{\rm L} W^+_\mu + {\rm h.c.} \; ,
\end{eqnarray}
where the CKM matrix $V$ appears in the form $V = O_{\rm u}^{\rm
T}P_{\rm u}^* P^{}_{\rm d}O^{}_{\rm d}$. The nine elements of $V$
can be explicitly expressed as
\begin{equation}\label{mixing}
  V^{}_{i\alpha} = O_{1i}^{\rm u}O_{1\alpha}^{\rm d}+O_{2i}^{\rm u}
  O_{2\alpha}^{\rm d}e^{{\rm i} \phi^{}_1}
  +O_{3i}^{\rm u}O_{3\alpha}^{\rm d}e^{{\rm i}
  \left(\phi^{}_1+\phi^{}_2\right)} \ ,
\end{equation}
where $\phi^{}_1 = \phi^{}_{C^{}_{\rm u}}-\phi^{}_{C^{}_{\rm d}}$
and $\phi^{}_2 = \phi^{}_{B^{}_{\rm u}}-\phi^{}_{B^{}_{\rm d}}$, and
the subscripts $i$ and $\alpha$ run over $(u,c,t)$ and $(d,s,b)$,
respectively. Now it is clear that $V$ depends on four free
parameters $A^{}_{\rm u}$, $A^{}_{\rm d}$, $\phi^{}_{1}$ and
$\phi^{}_{2}$, after the quark masses are input. With the help of
the above analytical results, we are able to constrain the parameter
space of $M^{}_{\rm u}$ and $M^{}_{\rm d}$ by taking account of the
latest values of the CKM matrix elements \cite{pdg}
\begin{equation}\label{ckm}
 |V|=\left(
 \begin{array}{ccc}
 0.97427\pm0.00014&\hspace{1cm} 0.22536 \pm 0.00061
 &\hspace{1cm}0.00355\pm{0.00015}\\
0.22522\pm0.00061 &\hspace{1cm} 0.97343 \pm 0.00015 &\hspace{1cm} 0.0414\pm{0.0012}\\
0.00886\pm{0.00033}&\hspace{1cm} 0.0405\pm{0.0012}&\hspace{1cm} 0.99914\pm{0.00005}
 \end{array}
 \right) \; ,
\end{equation}
together with the updated values of quark masses at the scale of
$M^{}_{Z}$ \cite{mass}
\begin{equation}\label{}
 \begin{array}{lll}
  m^{}_{u}=1.38^{+0.42}_{-0.41}\ {\rm MeV} \ ,
  &\hspace{1cm}m^{}_c=638^{+43}_{-84}\ {\rm MeV} \ ,
  &\hspace{1cm}m^{}_t=172.1\pm1.2\  {\rm GeV} \ , \\
  m^{}_d=2.82\pm 0.48\ {\rm MeV} \ ,
  &\hspace{1cm}m^{}_s=57^{+18}_{-12}\ {\rm MeV} \ ,
  &\hspace{1cm}m^{}_b=2860^{+160}_{-60}\ {\rm MeV} \ .
  \end{array}
\end{equation}
In our numerical analysis, we prefer to use $|V^{}_{us}|$,
$|V^{}_{ub}|$, $|V^{}_{cb}|$ and the CP-violating observable $\sin
2\beta$ as the inputs because their values have been determined to a
very good degree of accuracy. Here $\beta$ stands for one of the
inner angles of the CKM unitarity triangle described by the
orthogonality relation $V_{ub}^* V^{}_{ud} + V_{cb}^*V^{}_{cd} +
V_{tb}^*V^{}_{td}=0$ in the complex plane. The three inner angles of
this triangle are defined as
\begin{equation}\label{}
 \alpha=\arg\left(-\displaystyle
 \frac{V_{tb}^*V^{}_{td}}{V_{ub}^*V^{}_{ud}}
 \right)\ ,\hspace{1cm}
 \beta=\arg\left(-\displaystyle
 \frac{V_{cb}^*V^{}_{cd}}{V_{tb}^*V^{}_{td}}
 \right)\ ,\hspace{1cm}
 \gamma=\arg\left(-\displaystyle
 \frac{V_{ub}^*V^{}_{ud}}{V_{cb}^*V^{}_{cd}}
 \right)\ ,
\end{equation}
and their experimental values are \cite{pdg}
\begin{equation}\label{phase}
 \alpha=\left(85.4^{+3.9}_{-3.8}\right)^\circ\ ,\hspace{1cm}
 \sin{2\beta}=0.682\pm0.019\ ,\hspace{1cm}
 \gamma=\left(68.0^{+8.0}_{-8.5}\right)^\circ\ .
\end{equation}
Obviously, the uncertainty associated with $\sin 2\beta$ is much
smaller than those associated with $\alpha$ and $\gamma$. The
unitarity of $V$ requires $\alpha + \beta + \gamma = \pi$.
\begin{figure}
\begin{minipage}[t]{0.49\textwidth}
\includegraphics[width=3.2in]{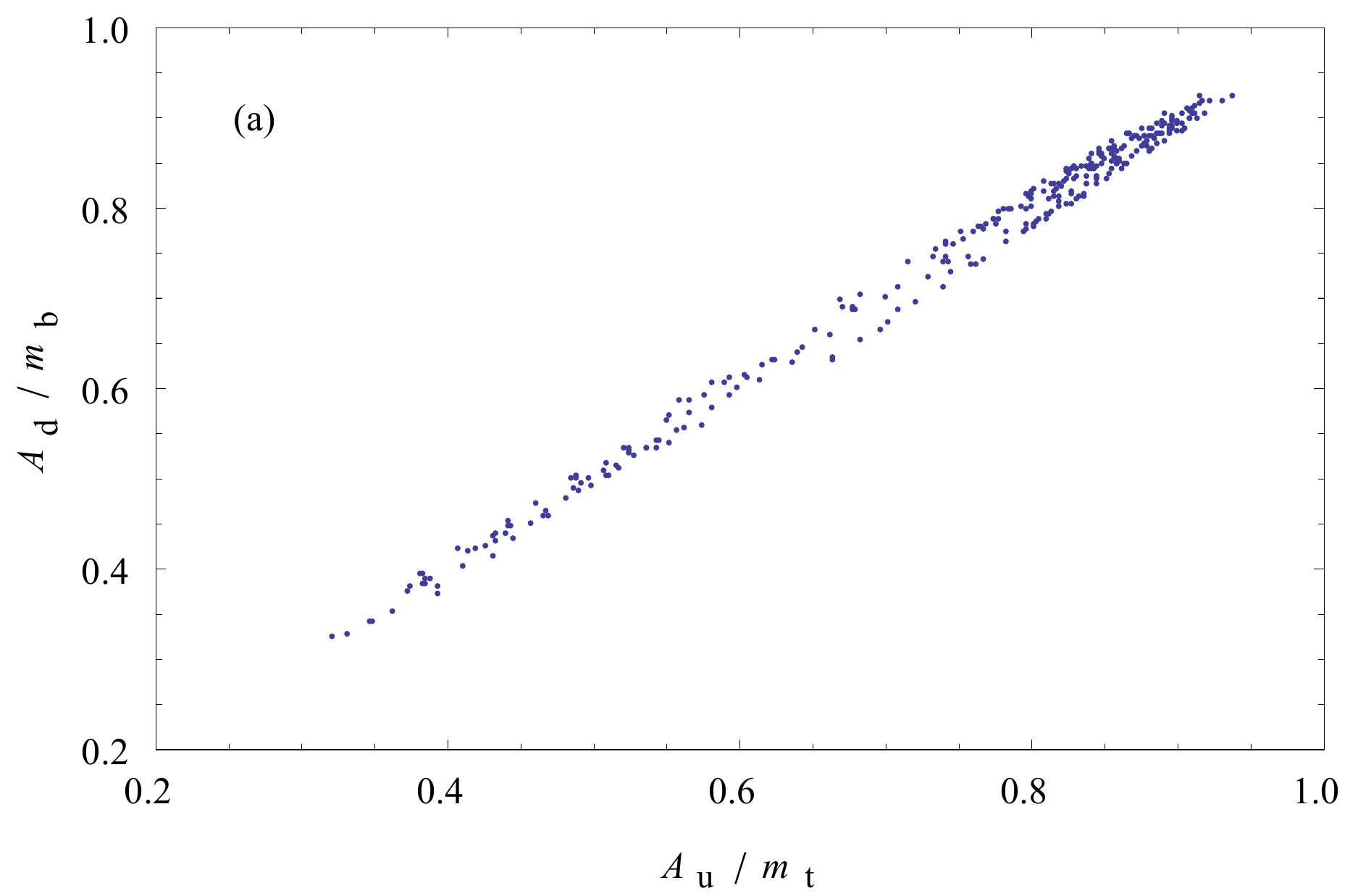}
\end{minipage}
\begin{minipage}[t]{0.49\textwidth}
\includegraphics[width=3.2in]{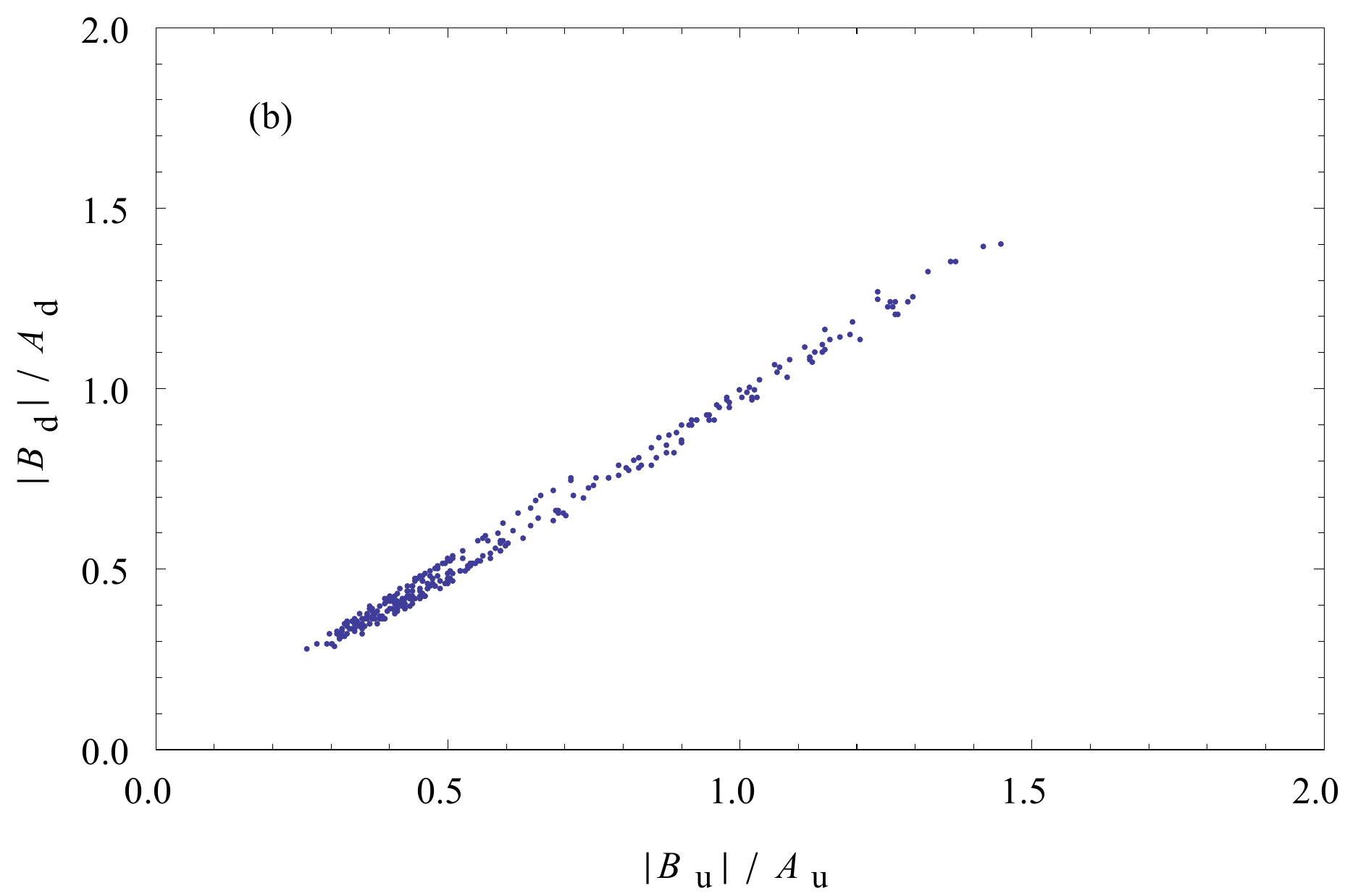}
\end{minipage}
\begin{minipage}[t]{0.49\textwidth}
\includegraphics[width=3.2in]{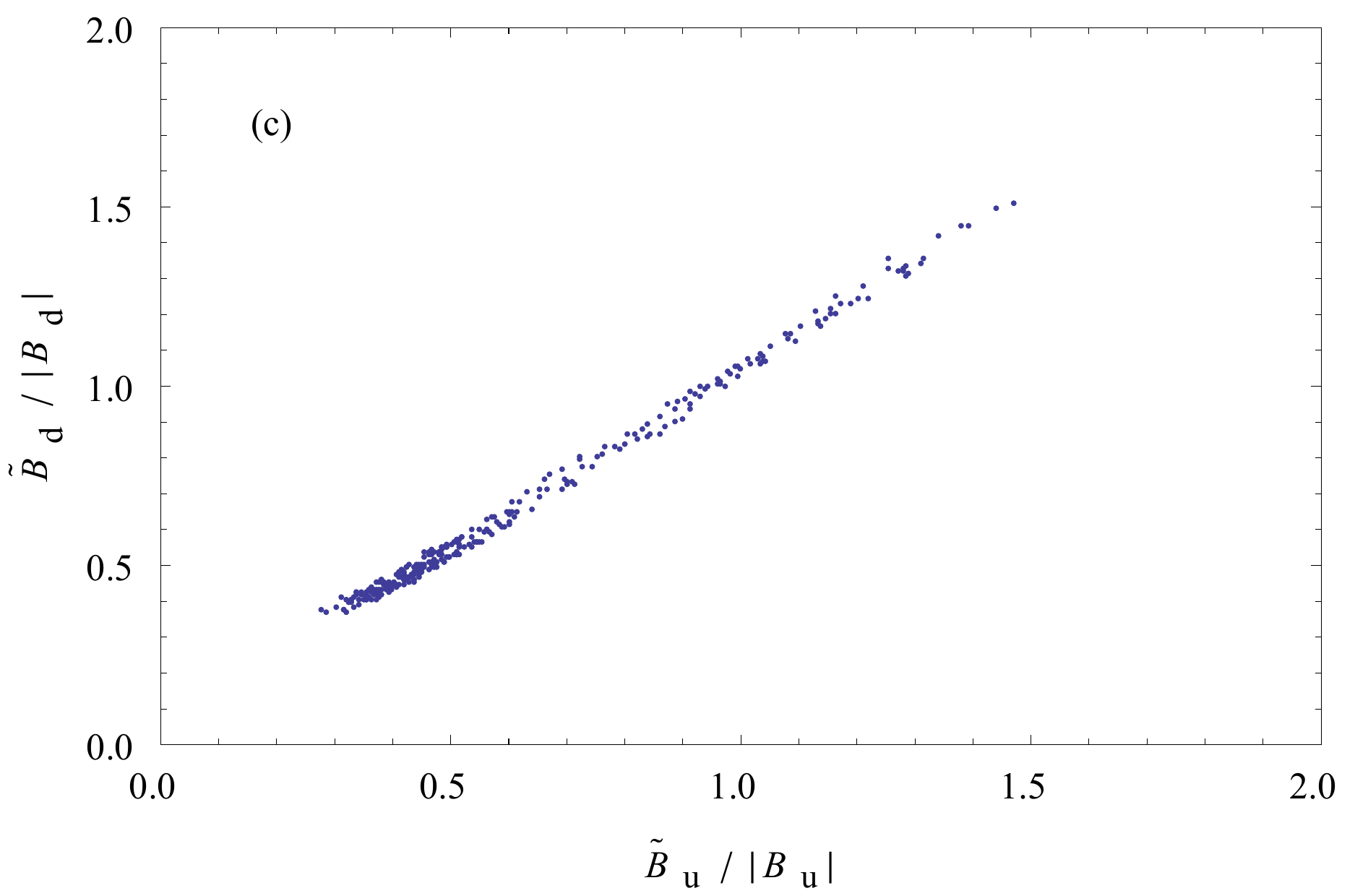}
\end{minipage}
\begin{minipage}[t]{0.49\textwidth}
\includegraphics[width=3.3in]{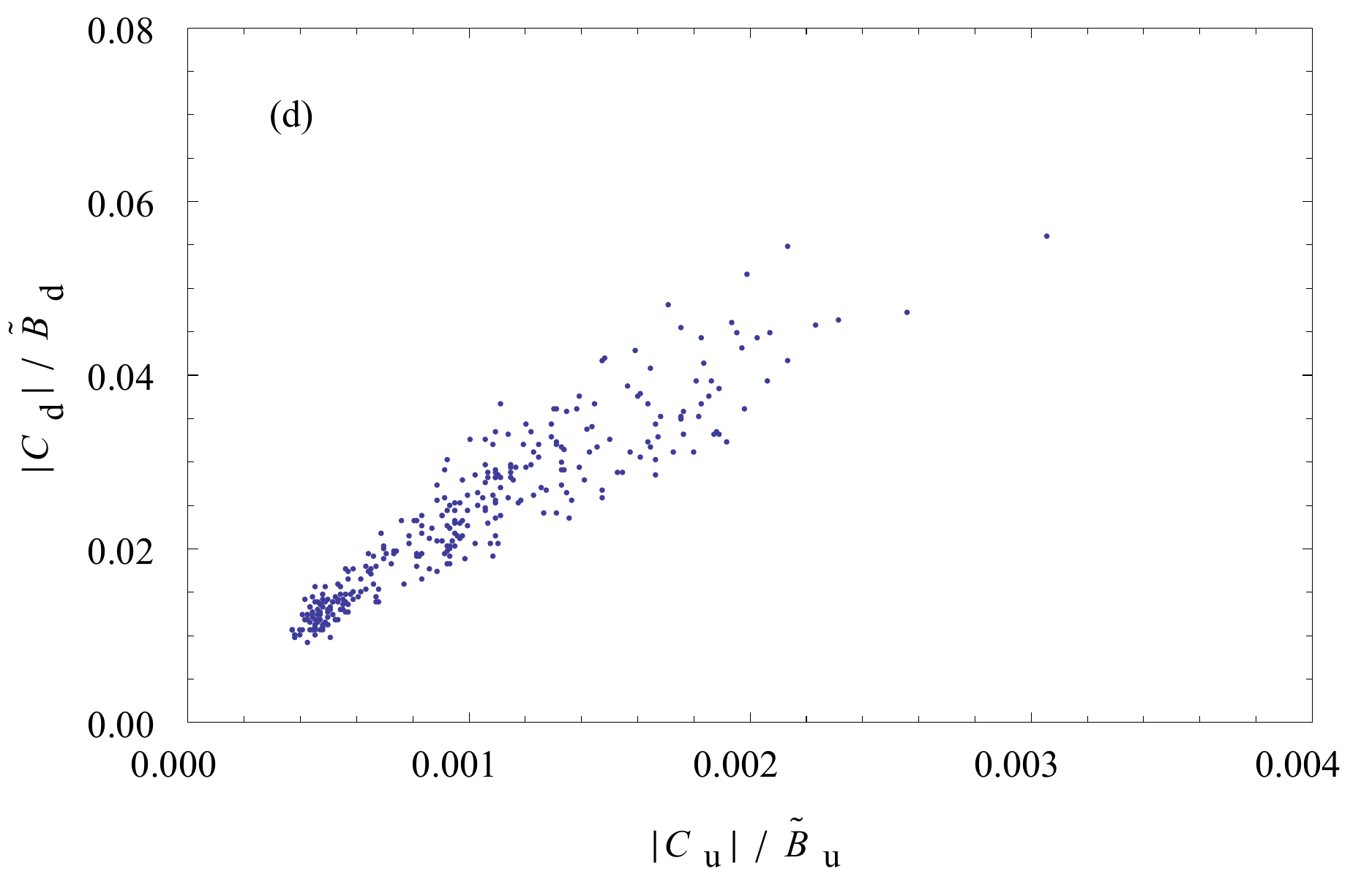}
\end{minipage}
\caption{The allowed regions of $A^{}_{\rm u,d}$, $|B^{}_{\rm
u,d}|$, $\tilde{B}^{}_{\rm u,d}$ and $|C^{}_{\rm u,d}|$ as
constrained by current experimental data in the $(\eta^{}_{\rm u},
\eta^{}_{\rm d}) =(+1, +1)$ case.}
 \label{fig1}
\end{figure}
\begin{figure}
\begin{minipage}[t]{0.49\textwidth}
\includegraphics[width=3.2in]{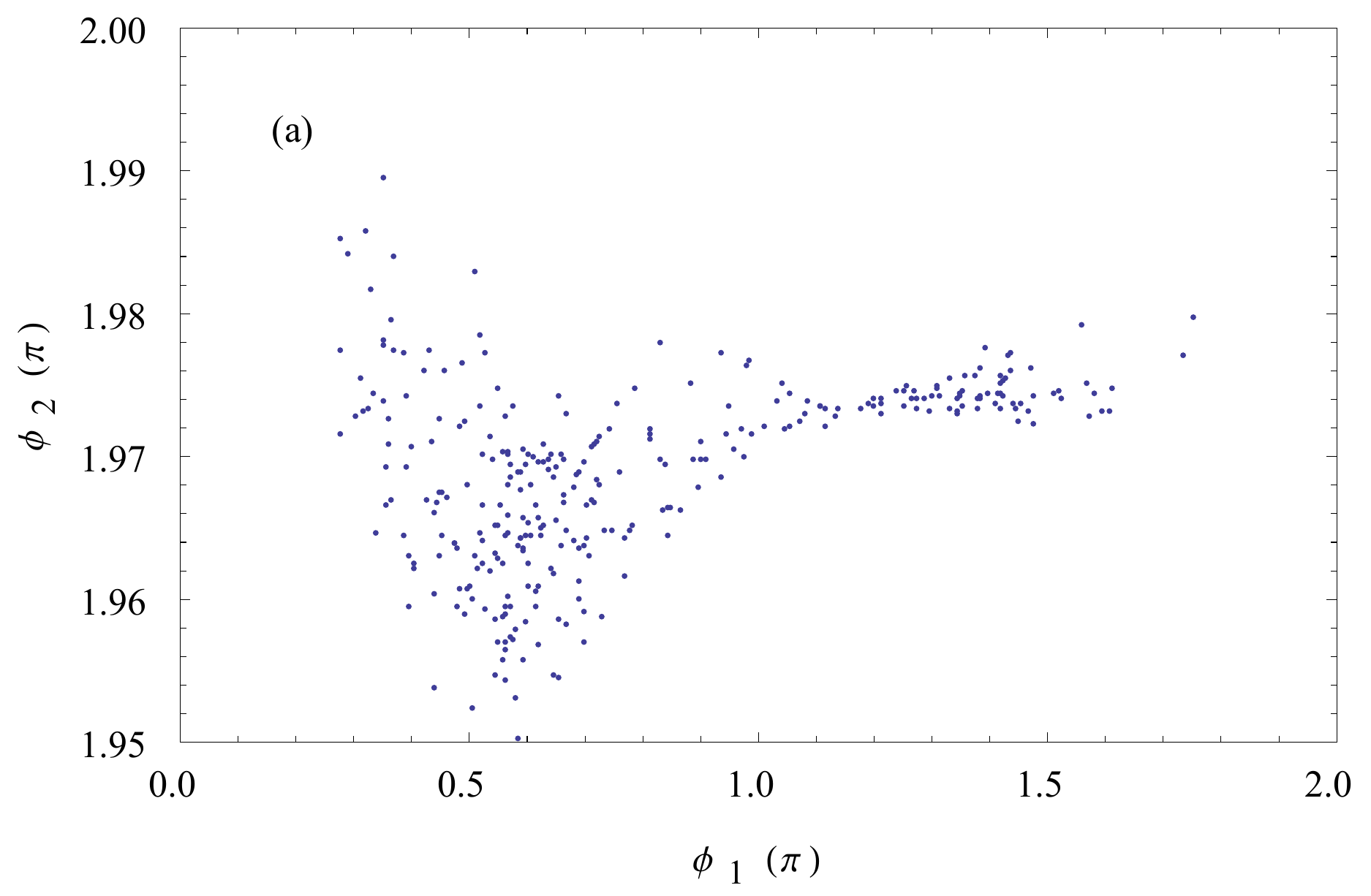}
\end{minipage}
\begin{minipage}[t]{0.49\textwidth}
\includegraphics[width=3.2in]{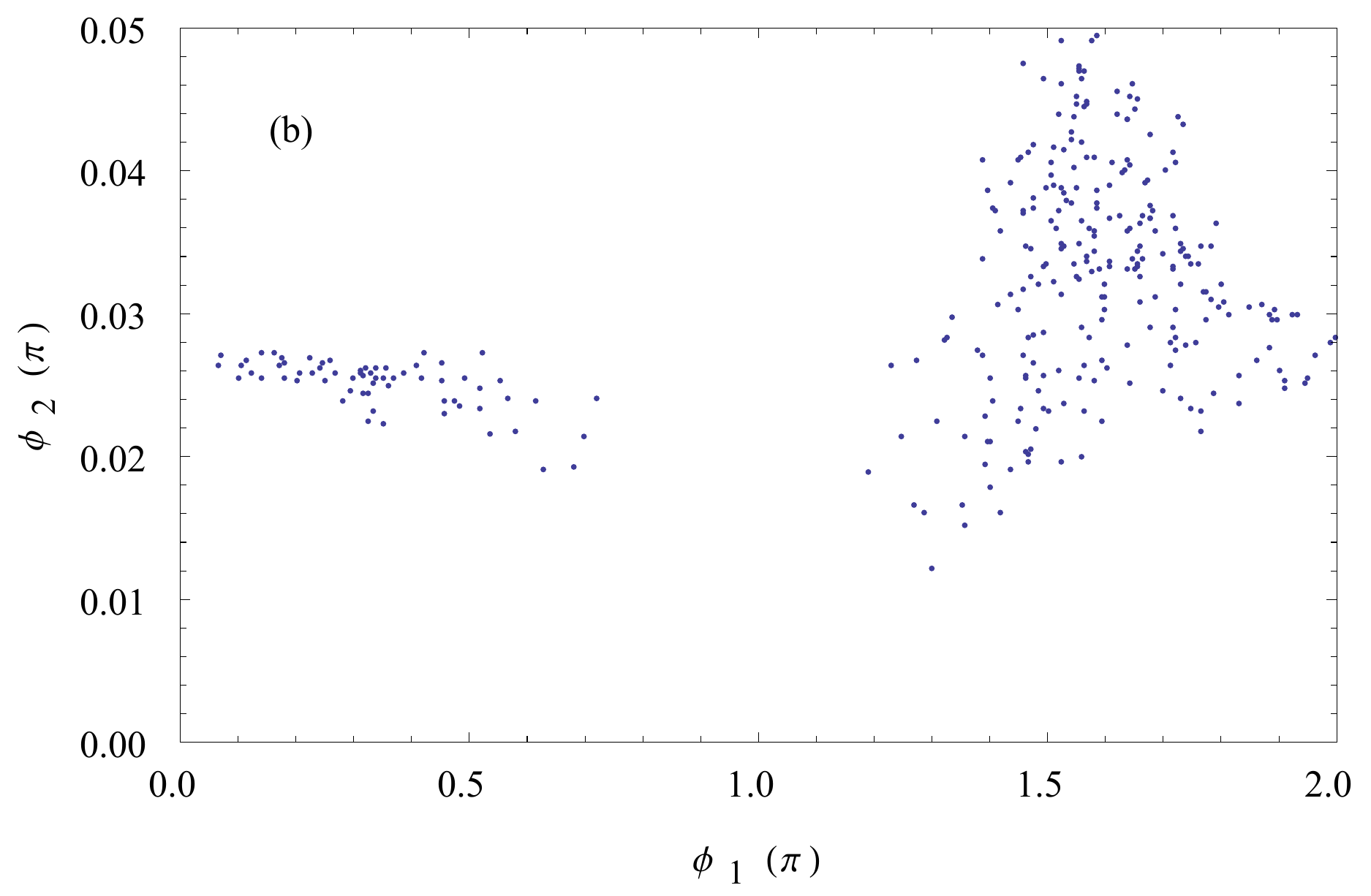}
\end{minipage}
\begin{minipage}[t]{0.49\textwidth}
\includegraphics[width=3.2in]{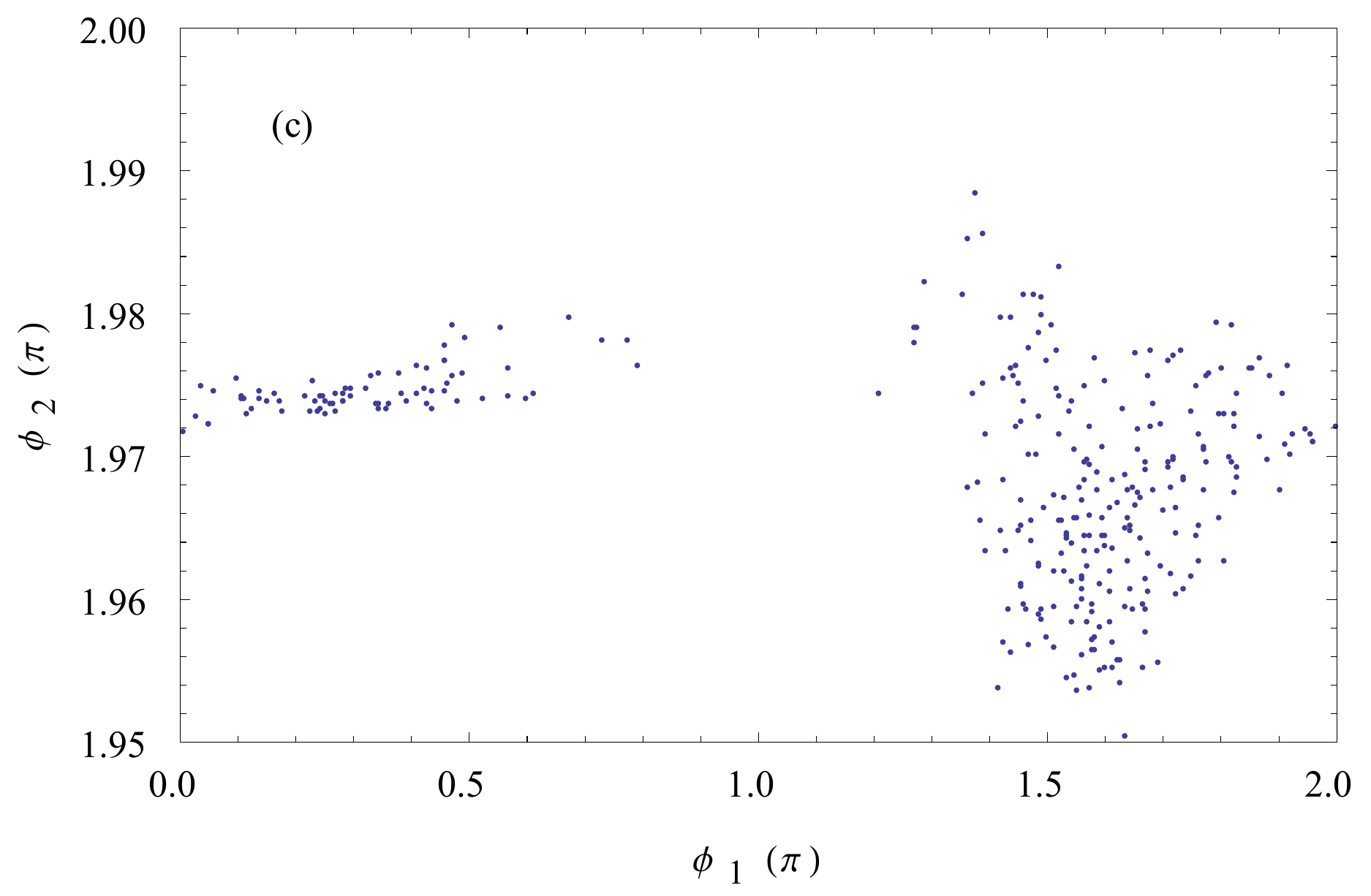}
\end{minipage}
\begin{minipage}[t]{0.49\textwidth}
\includegraphics[width=3.2in]{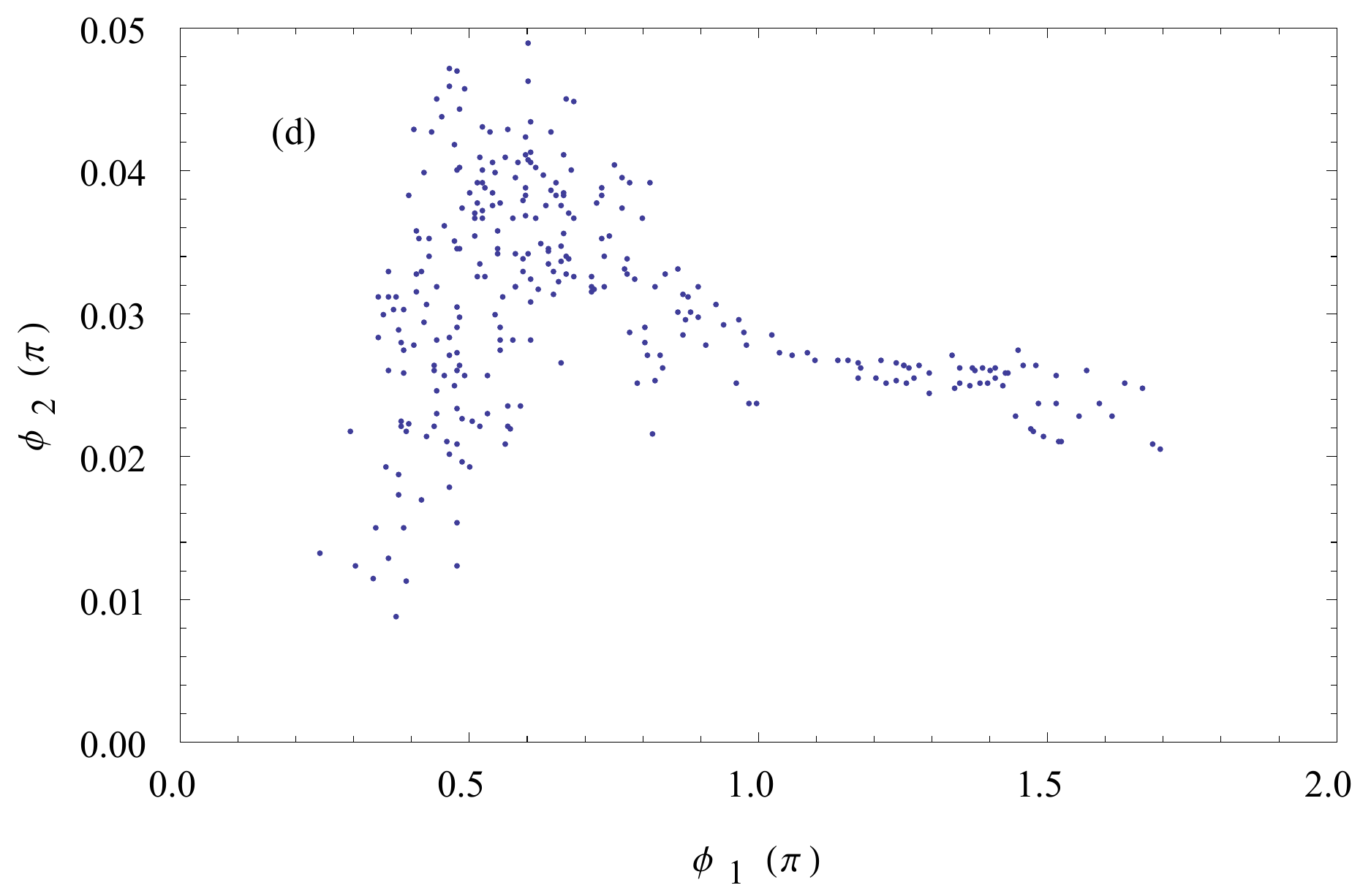}
\end{minipage}
\caption{The allowed regions of $\phi^{}_1$ and $\phi^{}_2$ as
constrained by current experimental data in the $(\eta^{}_{\rm u},
\eta^{}_{\rm d}) =(\pm 1, \pm1)$ cases.} \label{fig2}
\end{figure}

FIG. 1(a) shows the allowed region of $A^{}_{\rm u}$ and $A^{}_{\rm
d}$, which are rescaled as $r^{}_{\rm u} = A^{}_{\rm u}/m^{}_t$ and
$r^{}_{\rm d} = A^{}_{\rm d}/m^{}_b$, in the $(\eta^{}_{\rm u},
\eta^{}_{\rm d})=(+1, +1)$ case. Since the results of $r^{}_{\rm u}$
and $r^{}_{\rm d}$ in the other three cases are not quite different
from that illustrated in FIG. 1(a), here we just concentrate on the
$(\eta^{}_{\rm u}, \eta^{}_{\rm d})=(+1, +1)$ case for the sake of
simplicity. Now that $r^{}_{\rm u} \simeq r^{}_{\rm d}$ is a quite
good approximation as shown in FIG. 1(a), we simply use $r$ to
denote both $r^{}_{\rm u}$ and $r^{}_{\rm d}$ when their difference
needs not to be mentioned. We find that the region of $r$ can be
roughly divided into two parts: (1) $r$ is close to 1 and mainly
lies in the range of $0.8$ to $0.9$; (2) $r$ is around 0.5 and
mainly ranges from $0.4$ to $0.6$. These two parts will be referred
to as the $r \sim 1$ and $r \sim 0.5$ regions, respectively, in the
following discussions. The reasonableness of this treatment will
become clear shortly, since the phase parameters $\phi^{}_1$ and
$\phi^{}_2$ behave very differently in these two regions.

The allowed regions of $\phi^{}_1$ and $\phi^{}_2$ are shown in
FIG.~\ref{fig2}, where the possibilities of $(\eta^{}_{\rm u},
\eta^{}_{\rm d}) =(+1, +1), (+1, -1), (-1, +1)$ and $(-1, -1)$ have
all been considered. Taking the $(+1, +1)$ case for example, we find
that $\phi^{}_2$ is very close to 2$\pi$ and thus its allowed range
can also be denoted as $\phi^{}_2 \lesssim 0$. In comparison, the
allowed range of $\phi^{}_1$ is much wider but it can also be
divided into two parts in a reasonable approximation: $\phi^{}_1
\sim 0.5 \pi$ and $\phi^{}_1 \sim 1.5 \pi$. There is actually a
correlation between $r$ and $\phi^{}_1$: in the $r \sim 1$ region
$\phi^{}_1\sim 0.5 \pi$ holds, and in the $r \sim 0.5$ region
$\phi^{}_1\sim 1.5 \pi$ holds. After examining all the four
$(\eta^{}_{\rm u}, \eta^{}_{\rm d}) = (\pm 1, \pm 1)$ cases, we
obtain the more general correlation between $r$ and $\phi^{}_{1,2}$
as follows:
\begin{equation}\label{}
 \begin{array}{ll}
 \eta^{}_{\rm u}\ \eta^{}_{\rm d}\ \sin{\phi^{}_1}>0 \hspace{0.2cm}
 {\rm for} \hspace{0.2cm} r \sim 1\ ;&\hspace{0.7cm}
 \eta^{}_{\rm u}\ \eta^{}_{\rm d}\ \sin{\phi^{}_1}<0 \hspace{0.2cm}
 {\rm for} \hspace{0.2cm} r \sim 0.5\ ;\\
 \eta^{}_{\rm d}\ \sin{\phi^{}_2}<0\ ;&\hspace{0.7cm}  \eta^{}_{\rm u}\
 \eta^{}_{\rm d}\  \cos{\phi^{}_1}<0\ .\\
 \end{array}
\end{equation}
Note that only the constraint $ \eta^{}_{\rm d}\sin{\phi^{}_2}<0$ is
numerically exact, and the other three constraints serve for good
approximations in which most scattered points are satisfied. Such
correlative constraints can also be given in a more explicit way, as
listed in TABLE \Rmnum 1. Finally let us point out that the $r \sim
1$ region and its corresponding parameter correlation found here are
consistent with the results presented in Ref. \cite{xingzhang}, but
the $r \sim 0.5$ region and its parameter correlation are our new
findings which were missed in the previous works
(mainly because $\phi_1^{}$ takes totally different values in this region
from our expectation based on its values in the $r \sim 1$ region).
\begin{table}[t]
\caption{The correlation between $r$ and $\phi^{}_{1,2}$ in the four
$(\eta^{}_{\rm u}, \eta^{}_{\rm d}) = (\pm 1, \pm 1)$ cases.}

\begin{tabular}{|p{4cm}<{\centering}|p{6cm}<{\centering}|p{6cm}<{\centering}|}
\hline
&$r \sim 1$ &$ r \sim 0.5$ \\
         \hline
$(\eta^{}_{\rm u},\eta^{}_{\rm d})=(+1,+1)$ & $\phi^{}_1 \sim 0.5
\pi$\ , \hspace{0.1cm} $\cos{\phi^{}_1}<0$\ , \hspace{0.1cm}
$\phi^{}_2 \lesssim 0$\ . & $\phi^{}_1\sim  1.5 \pi$\
,\hspace{0.1cm} $\cos{\phi^{}_1}<0$\ ,\hspace{0.1cm}
$\phi^{}_2 \lesssim 0$\ .\\
         \hline
$(\eta^{}_{\rm u},\eta^{}_{\rm d})=(+1,-1)$ & $\phi^{}_1 \sim 1.5
\pi$\ , \hspace{0.1cm} $\cos{\phi^{}_1}>0$\ ,\hspace{0.1cm}
$\phi^{}_2 \gtrsim 0$\ . & $\phi^{}_1 \sim 0.5 \pi$\ ,\hspace{0.1cm}
$\cos{\phi^{}_1}>0$\ ,\hspace{0.1cm}
$\phi^{}_2 \gtrsim 0$\ .\\
         \hline
$(\eta^{}_{\rm u},\eta^{}_{\rm d})=(-1,+1)$ & $\phi^{}_1 \sim 1.5
\pi$\ , \hspace{0.1cm} $\cos{\phi^{}_1}>0$\ ,\hspace{0.1cm}
$\phi^{}_2 \lesssim 0$\ . & $\phi^{}_1\sim 0.5 \pi$\ ,\hspace{0.1cm}
$\cos{\phi^{}_1}>0$\ ,\hspace{0.1cm}
$\phi^{}_2 \lesssim 0$\ .\\
         \hline
$(\eta^{}_{\rm u},\eta^{}_{\rm d})=(-1,-1)$ & $\phi^{}_1 \sim 0.5
\pi$\ , \hspace{0.1cm} $\cos{\phi^{}_1}<0$\ ,\hspace{0.1cm}
$\phi^{}_2 \gtrsim 0$\ . & $\phi^{}_1 \sim 1.5 \pi$\ ,\hspace{0.1cm}
$\cos{\phi^{}_1}<0$\ ,
\hspace{0.1cm} $\phi^{}_2 \gtrsim 0$\ .\\
         \hline
\end{tabular}
\label{table}
\end{table}

All the correlative constraints listed in TABLE I can find an
explanation once the analytical expression of the CKM matrix $V$ is
explicitly presented. No matter whether the region $r \sim 1$ or $r
\sim 0.5$ is concerned, one can easily check that $A$ is close to
the mass of the third-family quark and thus it is much larger than
the masses of the first- and second-family quarks. As a result, the
orthogonal matrices $O^{}_{\rm u}$ and $O^{}_{\rm d}$ can approximate to
\begin{eqnarray}\label{}
O^{}_{\rm u} & \simeq & \left(
 \begin{array}{ccc}
 \vspace{0.3cm}
1&\hspace{0.5cm} \eta^{}_{\rm u}\sqrt{\displaystyle
\frac{m^{}_u}{m^{}_c}}&
\hspace{0.5cm} 0 \\
 \vspace{0.3cm}
 -\eta^{}_{\rm u}\sqrt{\displaystyle r^{}_{\rm u}\
 \frac{m^{}_u}{m^{}_c}}&\hspace{0.5cm}
 \sqrt{r^{}_{\rm u} }&\hspace{0.5cm} \sqrt{1-r^{}_{\rm u}} \\
  \vspace{0.3cm}
 \eta^{}_{\rm u}\sqrt{(1-r^{}_{\rm u})\
 \displaystyle\frac{m^{}_u}{m^{}_c}}&\hspace{0.5cm}
 -\sqrt{1-r^{}_{\rm u}}&\hspace{0.5cm}  \sqrt{r^{}_{\rm u}}\\
 \end{array}
 \right) \; , \nonumber \\
O^{}_{\rm d} & \simeq &
 \left(
 \begin{array}{ccc}
 \vspace{0.3cm}
1&\hspace{0.5cm} \eta^{}_{\rm d}\sqrt{\displaystyle
\frac{m^{}_d}{m^{}_s}}&\hspace{0.5cm}
\sqrt{\left(\displaystyle\frac{1}{r^{}_{\rm
d}}-1\right)\displaystyle
\frac{m^{}_d}{m^{}_b}\ \frac{m^{}_s}{m^{}_b}} \\
 \vspace{0.3cm}
 -\eta^{}_{\rm d}\sqrt{r^{}_{\rm d}\ \displaystyle
 \frac{m^{}_d}{m^{}_s}}&\hspace{0.5cm}
 \sqrt{r^{}_{\rm d} }&\hspace{0.5cm} \sqrt{1-r^{}_{\rm d}} \\
  \vspace{0.3cm}
 \eta^{}_{\rm d}\sqrt{(1-r^{}_{\rm d})\left(1-\  \displaystyle
 \frac{\eta^{}_{\rm d}}{r^{}_{\rm d}}
 \ \displaystyle\frac{m^{}_s}{m^{}_b}
 \right)\displaystyle\frac{m^{}_d}{m^{}_s}}
 &\hspace{0.5cm} -\sqrt{1-r^{}_{\rm d}}&\hspace{0.5cm}  \sqrt{r^{}_{\rm d}\
 \left(1-\  \displaystyle \frac{\eta^{}_{\rm d}}{r^{}_{\rm d}}
 \ \displaystyle\frac{m^{}_s}{m^{}_b} \right)}\\
 \end{array}
 \right) \; . \hspace{0.7cm}
\end{eqnarray}
Because $m^{}_u/m^{}_c \sim m^{}_c/m^{}_t \sim \sin^4\theta^{}_{\rm
C}$ and $m^{}_d/m^{}_s \sim m^{}_s/m^{}_b \sim \sin^2\theta^{}_{\rm
C}$ hold, the (1,3) entry of $O^{}_{\rm u}$ is negligibly small but
that of $O^{}_{\rm d}$ is not. Although the factor
$m^{}_s/(r^{}_{\rm d}m^{}_b)$ is actually much smaller than 1, it is
kept in the (3,1) and (3,3) entries of $O^{}_{\rm d}$ since it will
play a crucial role in explaining the correlation $\eta^{}_{\rm
d}\sin{\phi^{}_2}<0$.

Given the approximate results of $O^{}_{\rm u}$ and $O^{}_{\rm d}$
in Eq. (16), it is straightforward to calculate all the CKM matrix
elements by using Eq. (10). We are particularly interested in
\begin{eqnarray}\label{ckmelement}
\begin{aligned}
\vspace{0.5cm}
  |V^{}_{us}| \simeq &\left|\eta^{}_{\rm u}\ \eta^{}_{\rm d}
  \sqrt{\displaystyle\frac{m^{}_d}{m^{}_s}}
  -\sqrt{\displaystyle\frac{m^{}_u}{m^{}_c}}\ e^{{\rm i}\phi^{}_1}
  \left(\displaystyle\sqrt{r^{}_{\rm u}\ r^{}_{\rm d}}
 +\sqrt{\displaystyle(1-r^{}_{\rm u})\ (1-r^{}_{\rm d})}\ e^{{\rm i}\phi^{}_2}
 \right)\right| \; , \\
  \vspace{0.5cm}
    |V^{}_{cb}| \simeq &\left|\sqrt{\displaystyle r^{}_{\rm u}\ (1-r^{}_{\rm d})}
  -\sqrt{\displaystyle (1-r^{}_{\rm u})\ r^{}_{\rm d}\left(1-\  \displaystyle
  \frac{\eta^{}_{\rm d}}{r^{}_{\rm d}}
  \ \displaystyle\frac{m^{}_s}{m^{}_b} \right)}\ e^{{\rm i}\phi^{}_2}\right| \; ,
  \\
  \vspace{0.5cm}
    {|V^{}_{ub}|} \simeq &\left|\sqrt{\displaystyle \frac{m^{}_d}{m^{}_b}\
    \frac{m^{}_s}{m^{}_b}\left(\displaystyle \frac{1}{r^{}_{\rm d}}-1\right)}
    -\eta^{}_{\rm u} \sqrt{\displaystyle\frac{m^{}_u}{m^{}_c}}\
    V^{}_{cb}\right| \; .\\
  \end{aligned}
\end{eqnarray}
Among them $|V^{}_{cb}|$ deserves special attention and can be
decomposed as follows:
\begin{equation}\label{vcb}
\begin{array}{ll}
\vspace{0.3cm}
&  |V^{}_{cb}|=\left|\ {\rm Re}\ (V_{cb}^\prime)- {\rm i}
\ {\rm Im}\ (V_{cb}^\prime)\ \right|
  =\sqrt{\left[{\rm Re}(V_{cb}^\prime)\right]^2
  + \left[{\rm Im}(V_{cb}^\prime)\right]^2}\ ,\\
\vspace{0.3cm}
&  {\rm Re}(V_{cb}^\prime) \simeq \sqrt{r^{}_{\rm u}\
(1-r^{}_{\rm d})}
  -\sqrt{(1-r^{}_{\rm u})\ r^{}_{\rm d}\ \left(1-\  \displaystyle
  \frac{\eta^{}_{\rm d}}{r^{}_{\rm d}}
  \ \displaystyle\frac{m^{}_s}{m^{}_b} \right)}\ \cos{\phi^{}_2}\ ,\\
&  {\rm Im}(V_{cb}^\prime) \simeq \sqrt{(1-r^{}_{\rm u})\ r^{}_{\rm d}\
\left(1-\ \displaystyle \frac{\eta^{}_{\rm d}}{r^{}_{\rm d}}
  \ \displaystyle\frac{m^{}_s}{m^{}_b} \right)}\ \sin{\phi^{}_2}\ ,
  \end{array}
\end{equation}
where $V_{cb}^\prime = e^{-{\rm i} \phi^{}_1} V^{}_{cb}$ has been
defined. Clearly, neither $|{\rm Re}\ (V_{cb}^\prime)|$ nor $|{\rm
Im}\ (V_{cb}^\prime)|$ is allowed to be larger than the experimental
result $|V^{}_{cb}|\simeq 0.04$. That is why $r^{}_{\rm u}$ is
always nearly equal to $r^{}_{\rm d}$ and $\phi^{}_2$ is so close to
$0$ or $2\pi$. For either $r^{}_{\rm u} \sim r^{}_{\rm d} \sim 1$ or
$r^{}_{\rm u} \sim r^{}_{\rm d} \sim 0.5$, the fact of $\phi^{}_2
\sim 0$ (or $2\pi$) allows us to simplify the expression of
$|V^{}_{us}|$ to
\begin{equation}\label{}
\begin{array}{l}
  |V^{}_{us}| \simeq \left|\eta^{}_{\rm u}\ \eta^{}_{\rm d}
  \sqrt{\displaystyle
  \frac{m^{}_d}{m^{}_s}}-
  \sqrt{\displaystyle\frac{m^{}_u}{m^{}_c}}\ e^{{\rm i}\phi^{}_1}\right|
  =\sqrt{\displaystyle\frac{m^{}_d}{m^{}_s} -
  2\ \eta^{}_{\rm u}\ \eta^{}_{\rm d}\
  \sqrt{\displaystyle\frac{m^{}_u}{m^{}_c}\
  \displaystyle\frac{m^{}_d}{m^{}_s} }
  \ \cos{\phi^{}_1}+\displaystyle\frac{m^{}_u}{m^{}_c}} \ .
\end{array}
\end{equation}
It is known that the term $\sqrt{m^{}_d/m^{}_s}$ itself can fit the
experimental value of $|V^{}_{us}|$ to a good degree of accuracy
(i.e., the GST relation), and hence one has to control the
contribution from the smaller term $\sqrt{m^{}_u/m^{}_c}$ by
adjusting the CP-violating phase $\phi^{}_1$. This observation
immediately leads to $\cos{\phi^{}_1} \sim 0$, or equivalently
$\phi^{}_1 \sim 0.5\pi$ or $1.5\pi$. As first pointed out in Ref.
\cite{FX95}, the relation in Eq. (19) is essentially compatible with
the orthogonality relation $V_{ub}^* V^{}_{ud} + V_{cb}^*V^{}_{cd} +
V_{tb}^*V^{}_{td}=0$ after the latter is rescaled by $V^*_{cb}$,
leading to the striking prediction $\alpha \simeq \phi^{}_1 \sim
0.5\pi$ for the corresponding CKM unitarity triangle. Needless to
say, this prediction is consistent with current experimental data
shown in Eq. (14).

In order to understand the correlation between the signs of
$\sin\phi^{}_{1,2}$ and those of $\eta^{}_{\rm u, d}$, one needs to
consider the impact of the CP-violating observable $\sin 2\beta$ on
the parameter space of $M^{}_{\rm u}$ and $M^{}_{\rm d}$. Eqs. (10)
and (16) allow us to obtain
\begin{equation}\label{}
\begin{array}{lll}
\vspace{0.2cm} {\rm Re} (V^{}_{cd}V_{cb}^*)& \simeq &-\eta^{}_{\rm
u}\ \displaystyle \sqrt{ \frac{m^{}_u}{m^{}_c}}\  \sin{\phi^{}_1}\
{\rm Im}(V_{cb}^\prime) + \left[\eta^{}_{\rm u}\ \displaystyle
\sqrt{ \frac{m^{}_u}{m^{}_c}}\cos{\phi^{}_1}-\eta^{}_{\rm d}\
\displaystyle \sqrt{ \frac{m^{}_d}{m^{}_s}}\ \right] {\rm
Re}(V_{cb}^\prime) \ ,\\ \vspace{0.2cm} {\rm Im}
(V^{}_{cd}V_{cb}^*)& \simeq &-\eta^{}_{\rm u}\ \displaystyle \sqrt{
\frac{m^{}_u}{m^{}_c}}\ \sin{\phi^{}_1}\ {\rm Re}(V_{cb}^\prime)
-\left[\eta^{}_{\rm u}\ \displaystyle \sqrt{
\frac{m^{}_u}{m^{}_c}}\cos{\phi^{}_1}-\eta^{}_{\rm d} \
\displaystyle \sqrt{ \frac{m^{}_d}{m^{}_s}}\ \right]
{\rm Im}(V_{cb}^\prime) \ ,\\
\vspace{0.2cm} {\rm Re} (V^{}_{td}V_{tb}^*)& \simeq &-\eta^{}_{\rm
d}\ \displaystyle \sqrt{ \frac{m^{}_d}{m^{}_s}}\ \left[
\displaystyle \frac{\eta^{}_{\rm d}}{r^{}_{\rm d}}\
\displaystyle\frac{m^{}_s}{m^{}_b} \sqrt{r^{}_{\rm
u}\left(1-r^{}_{\rm d}\right )}
-{\rm Re}(V_{cb}^\prime) \right] \ ,\\
{\rm Im} (V^{}_{td}V_{tb}^*)& \simeq &-\eta^{}_{\rm d}\
\displaystyle \sqrt{ \frac{m^{}_d}{m^{}_s}}\ {\rm Im}(V_{cb}^\prime) \ .
\end{array}
\end{equation}
Then the definition of $\beta$ in Eq. (13) leads us to
\begin{eqnarray}\label{beta}
\tan\beta & = & -\frac{{\rm Re} (V^{}_{cd}V_{cb}^*)\ {\rm Im}
(V^{}_{td}V_{tb}^*)
  - {\rm Im} (V^{}_{cd}V_{cb}^*)\ {\rm Re} (V^{}_{td}V_{tb}^*)}
{{\rm Re} (V^{}_{cd}V_{cb}^*)\ {\rm Re} (V^{}_{td}V_{tb}^*)
  + {\rm Im} (V^{}_{cd}V_{cb}^*)\ {\rm Im} (V^{}_{td}V_{tb}^*)}
  \hspace{0.5cm} \nonumber \\
& \simeq & \eta^{}_{\rm u}\ \eta^{}_{\rm d}\ \sin{\phi^{}_1}\
\displaystyle \sqrt{\frac{m^{}_u}{m^{}_c}\ \frac{m^{}_s}{m^{}_d}} \
- \ \displaystyle \frac{\eta^{}_{\rm d}}{r^{}_{\rm d}}\
\frac{m^{}_s}{m^{}_b} \ \sqrt{r^{}_{\rm u}\left(1-r^{}_{\rm d}\right)}
\nonumber \\
& & \times \left[1-\eta^{}_{\rm u}\ \eta^{}_{\rm d}\
\cos{\phi^{}_1}\sqrt{\displaystyle\frac{m^{}_u}{m^{}_c}\
\frac{m^{}_s}{m^{}_d}}\right] \displaystyle
\frac{ {\rm Im}(V_{cb}^\prime)}{|V^{}_{cb}|^2} \ .
\end{eqnarray}
Given the experimental value of $\sin 2\beta$ in Eq. (14), we arrive
at $\tan\beta = 0.394 \pm 0.015$. In the $r \sim 1$ region the first
term of Eq. (21) is dominant, and thus $\eta^{}_{\rm u}\eta^{}_{\rm
d}\sin{\phi^{}_1}$ is required to be positive. Note that this term
is at most 0.322, if the values of quark masses in Eq. (12) are
input. Hence the second term of Eq. (21) has to be positive too. In
other words, $\eta^{}_{\rm d}\sin{\phi^{}_2}$ should be negative
because ${\rm Im}(V_{cb}^\prime)$ is proportional to
$\sin{\phi^{}_2}$. Furthermore, $\eta^{}_{\rm u}\eta^{}_{\rm
d}\cos{\phi^{}_1}$ is likely to be negative to enhance the
contribution of the second term of Eq. (21) to $\tan\beta$. When the
$r \sim 0.5$ region is concerned, we find that the second term of
Eq. (21) becomes important, so $\eta^{}_{\rm d}\sin{\phi^{}_2}$ is
still required to be negative. Since this term has a chance to
saturate the experimental value of $\tan{\beta}$, the first term of
Eq. (21) is possible to be negative in such a case. In fact,
$\eta^{}_{\rm u}\eta^{}_{\rm d}\sin{\phi^{}_1}$ must be negative in
the $r \sim 0.5$ region if we take into account the constraint from
$|V^{}_{ub}|$. With the help of Eqs. (17) and (18), we have
\begin{equation}\label{vub}
{|V^{}_{ub}|} \simeq \left|\sqrt{\displaystyle
\frac{m^{}_d}{m^{}_b}\hspace{0.1cm} \frac{m^{}_s}{m^{}_b}
\left(\displaystyle \frac{1}{r^{}_{\rm d}}-1\right)}
   + \eta^{}_{\rm u}\ \eta^{}_{\rm d}\ \sin{\phi^{}_1}\
   \sqrt{\displaystyle\frac{m^{}_u}{m^{}_c}}\hspace{0.1cm}
   \left|{\rm Im}(V_{cb}^\prime) \right|\right| \; .
\end{equation}
Taking $r^{}_{\rm d}=0.5$ for example, we find that the first term of
Eq. (22) is about 0.0044, larger than the experimental value
$|V_{ub}| \simeq 0.0036$. In this case the second term of Eq. (22)
should be negative so as to partially offset the contribution from
the first term. Therefore, we are left with
$\eta^{}_{\rm u}\eta^{}_{\rm d}\sin{\phi^{}_1} <0$.
However, when the first term is large enough, the second term will
fail to offset its extra contribution, bringing about a lower bound
about 0.3 for $r_{\rm d}^{}$ as shown in FIG. 1(a).
\begin{figure}
\begin{minipage}[t]{0.49\textwidth}
\includegraphics[width=3in]{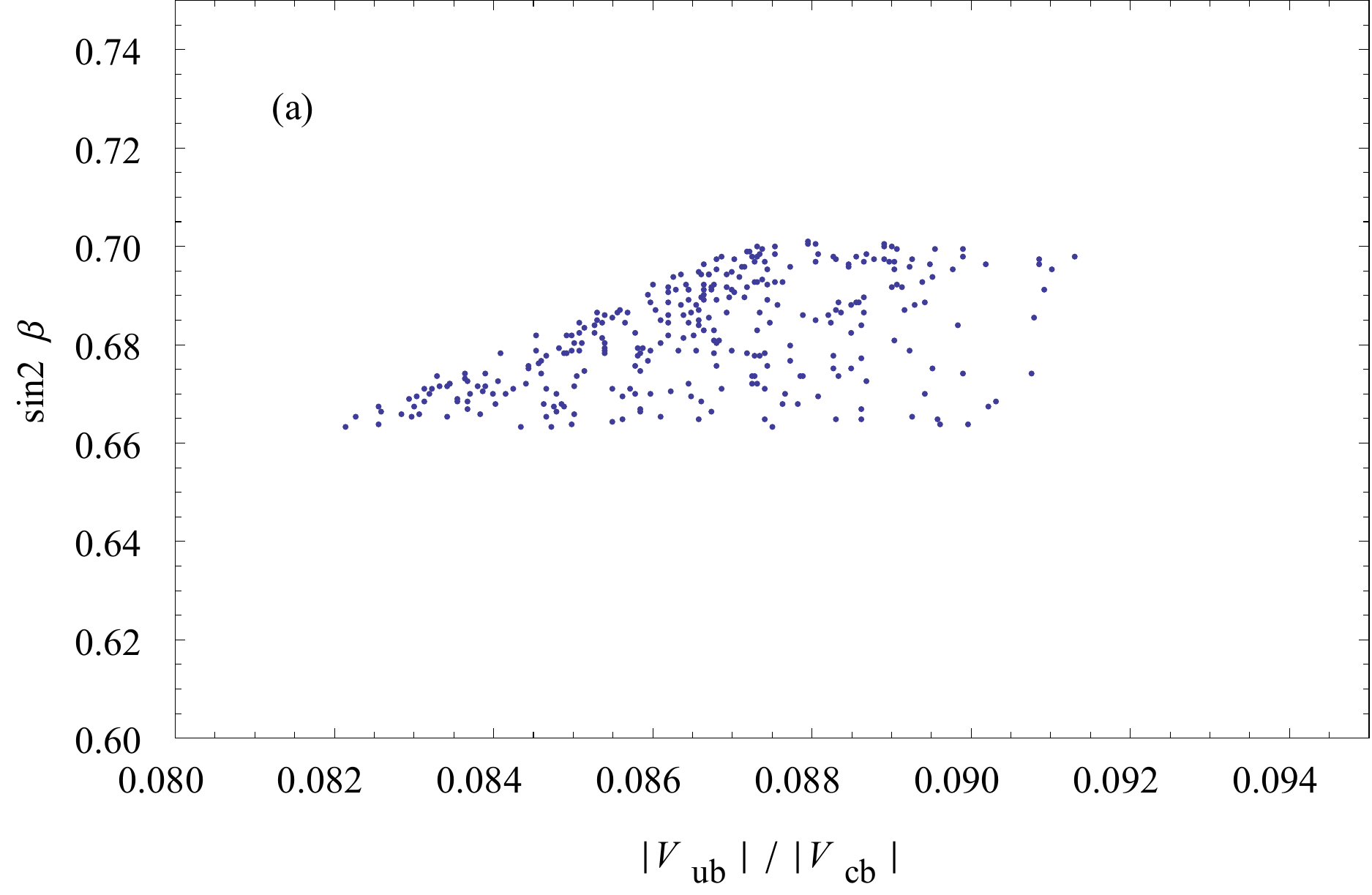}
\end{minipage}
\begin{minipage}[t]{0.49\textwidth}
\includegraphics[width=3in]{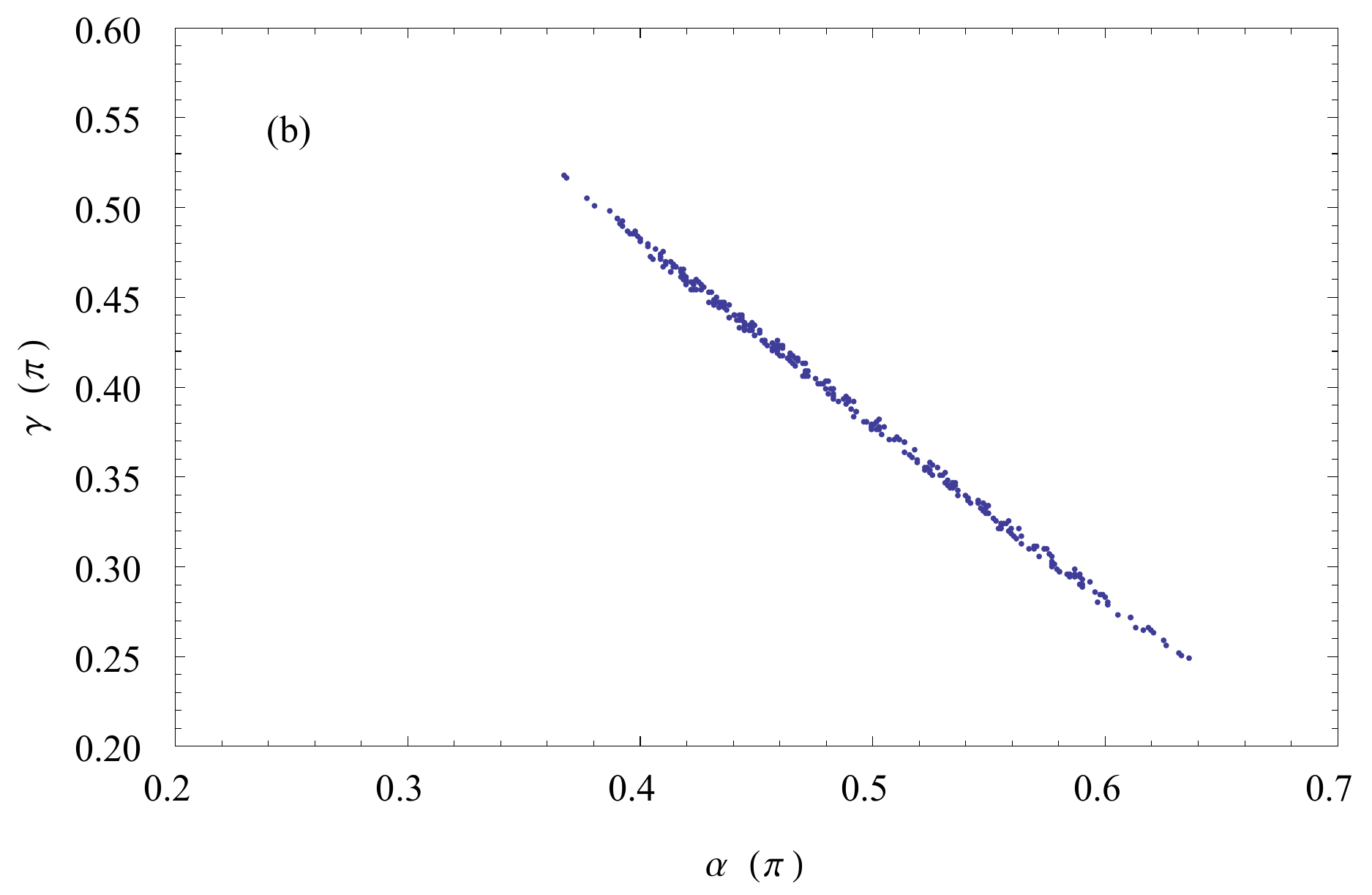}
\end{minipage}
\caption{The numerical outputs of $|V^{}_{\rm ub}|/|V^{}_{\rm cb}|$
versus $\sin{2\beta}$ and $\alpha$ versus $\gamma$ in the
$(\eta^{}_{\rm u}, \eta^{}_{\rm d})=(+1, +1)$ case.} \label{fig3}
\end{figure}

To summarize, we have performed a new numerical analysis of the
four-zero ansatz of quark mass matrices by using the updated values
of quark masses and CKM parameters. We find a new part of the
parameter space of this ansatz --- the $r\sim0.5$ region together
with the relevant correlation between $\phi^{}_{1, 2}$ and
$\eta^{}_{\rm u, d}$. We have also explained the salient features of
the whole parameter space of $M^{}_{\rm u}$ and $M^{}_{\rm d}$ in
some analytical approximations. As a byproduct, FIG.~\ref{fig3}
shows the numerical outputs of $|V^{}_{\rm ub}|/|V^{}_{\rm cb}|$
versus $\sin{2\beta}$ and $\alpha$ versus $\gamma$ in the
$(\eta^{}_{\rm u}, \eta^{}_{\rm d})=(+1, +1)$ case. One can see that
the uncertainties associated with the CP-violating quantities
$\alpha$ and $\gamma$ remain quite significant, and they mainly
originate from the uncertainties of $m^{}_s$, $r^{}_{\rm u}$ and
$r^{}_{\rm d}$. In the next section, we shall go back to the quark
mass matrices themselves to look at their structures and to see
whether they assume some special patterns with fewer free
parameters.

\section{Special four-zero patterns and model building}

In the Fritzsch ansatz of quark mass matrices, the first term of
Eq.~(\ref{vub}) should be replaced with $\left(m^{}_s/m^{}_b\right)
\sqrt{m^{}_d/m^{}_b}~$, whose size is about 0.00078. This result can
easily be understood from the trace of $M^{\rm F}_{\rm u, d}$ (i.e.,
$A=\lambda^{}_1+\lambda^{}_2+\lambda^{}_3$), which gives rise to
$r^{}_{\rm d}=1-\mathcal O(m^{}_s/m^{}_b)$ in the down sector. Since
the upper limit that the second term of Eq.~(\ref{vub}) can reach is
about 0.00236, the experimental value of $|V^{}_{ub}|$ has no way to
be saturated by these two terms in the Fritzsch ansatz. When the
four-zero texture of $M^{}_{\rm u,d}$ is concerned, the existence of
nonzero (2,2) entries modifies the trace of $M^{}_{\rm u,d}$ to the
form $A+\tilde B = \lambda^{}_1+\lambda^{}_2+\lambda^{}_3$. The
constraint on $r^{}_{\rm d}$ is consequently relaxed, and it
actually becomes a free parameter. Given a typical value $r^{}_{\rm
d}=0.9$ in the $r\sim 1$ region, for example, the first term of
Eq.~(\ref{vub}) contributes a value 0.0014 to $|V^{}_{ub}|$ such
that the experimental result of $|V^{}_{ub}|$ can be well fitted. In
this case the magnitude of $\tilde B$ is about 0.1$\lambda^{}_3$. To
avoid a relatively large $\lambda^{}_2$ from such a large $\tilde
B$, the three parameters $A$, $\tilde B$ and $|B|$ must satisfy an
approximate geometrical relation \cite{fourzero2} up to a correction
of $\mathcal O(m^{}_2/m^{}_3)$:
\begin{equation}\label{abtb}
\displaystyle \frac{|B|}{A} \simeq \displaystyle \frac{\tilde B}{|B|}
\left[1+\mathcal O\left(\displaystyle \frac{m^{}_2}{m^{}_3}
\right)\right]\ .
\end{equation}
This observation is certainly supported by the numerical results
presented in FIG.~\ref{fig1}. In light of the definition $A/m^{}_3 =
r$, $\tilde B/m^{}_3 \simeq 1-r$ holds as a good approximation
because of $|\lambda^{}_1 + \lambda^{}_2| \ll \lambda^{}_3 =
m^{}_3$. Hence Eq.~(\ref{abtb}) implies $|B|/m^{}_3 \simeq \sqrt{r
\left(1-r\right)}~$. In short, the (2,3) sectors of $\overline
M^{}_{\rm u}$ and $\overline M^{}_{\rm d}$ have the same structure
which can be parameterized as
\begin{equation}\label{2309}
\overline M_{\rm u}^{(2,3)}\sim
 A^{}_{\rm u}\left(
 \begin{array}{cc}
\epsilon^2 &\hspace{0.3cm}\epsilon \\
 \epsilon &\hspace{0.3cm}1
 \end{array}
 \right) \; ,
 \hspace{1cm}
 \overline M_{\rm d}^{(2,3)}\sim
 A^{}_{\rm d}\left(
 \begin{array}{cc}
\epsilon^2 &\hspace{0.3cm}\epsilon \\
 \epsilon &\hspace{0.3cm}1
 \end{array}
 \right) \; ,
\end{equation}
where $\epsilon \simeq \sqrt{\left(1-r\right)/r}~$, and its value is
about 0.3 in the $r \sim 1$ region of the parameter space.
Eq.~(\ref{2309}) hints at a common origin of the (2,3) sectors of
$\overline M^{}_{\rm u}$ and $\overline M^{}_{\rm d}$, and thus it
can be taken as a guideline for model building. However, a numerical
analysis shows that such an up-down parallelism is slightly broken
by the (2,2) entries of quark mass matrices. In addition, their
(1,2) entries do not share this kind of parallelism, as one can see
in FIG.~\ref{fig1}(d). With the help of Eq.~(\ref{parameter}), we
typically take $r^{}_{\rm u} \simeq r^{}_{\rm d} \simeq 0.9$ and
illustrate the finite matrix elements of $\overline M^{}_{\rm u}$
and $\overline M^{}_{\rm d}$ as follows:
\begin{equation}\label{r09}
\overline M^{}_{\rm u} \simeq
 A^{}_{\rm u}\left(
 \begin{array}{ccc}
0 &\hspace{0.2cm}0.0002 &\hspace{0.2cm}0\\
 0.0002 &\hspace{0.2cm}0.11&\hspace{0.2cm}0.31\\
0 &\hspace{0.2cm}0.31 &\hspace{0.2cm}1
 \end{array}
 \right) \; ,
 \hspace{1cm}
 \overline M^{}_{\rm d} \simeq
 A^{}_{\rm d}\left(
 \begin{array}{ccc}
0 &\hspace{0.2cm}0.005 &\hspace{0.2cm} 0\\
 0.005 &\hspace{0.2cm}0.13 &\hspace{0.2cm}0.31\\
 0&\hspace{0.2cm}0.31 &\hspace{0.2cm}1
 \end{array}
 \right) \; .
\end{equation}
It is worth reiterating that the mild hierarchy in the (2,3) sectors
of quark mass matrices is crucial to fit current experimental data.

In the $r\sim 0.5$ region of the parameter space of quark mass matrices
$M^{}_{\rm u}$ and $M^{}_{\rm d}$, there is a particularly
interesting case,
\begin{equation}\label{}
  A^{}_{\rm u}=\tilde B^{}_{\rm u}\ , \hspace{1cm}
  A^{}_{\rm d}=\tilde B^{}_{\rm d}\ ,
\end{equation}
which deserves special attention. We have verified that
these exact equalities are really allowed in our numerical calculations.
The corresponding parameter space is certainly a part of the parameter
space restricted by $r\sim 0.5$. In this special case, the (2,3) sectors
of $\overline M^{}_{\rm u}$ and $\overline M^{}_{\rm d}$ have a neat form:
\begin{equation}\label{2305}
\overline M_{\rm u}^{(2,3)}\sim
 A^{}_{\rm u}\left(
 \begin{array}{cc}
 \vspace{0.2cm}
1 &\hspace{0.2cm}1-  \displaystyle \frac{2m^{}_c}{m^{}_t} \\
 1-  \displaystyle \frac{2m^{}_c}{m^{}_t} &\hspace{0.2cm}1
 \end{array}
 \right) \; ,
 \hspace{1cm}
 \overline M_{\rm d}^{(2,3)}\sim
 A^{}_{\rm d}\left(
 \begin{array}{cc}
 \vspace{0.2cm}
1 &\hspace{0.2cm}1- \displaystyle \frac{2m^{}_s}{m^{}_b}  \\
 1-  \displaystyle \frac{2m^{}_s}{m^{}_b}  &\hspace{0.2cm}1
 \end{array}
 \right) \; .
\end{equation}
A typical numerical illustration of the structures of $\overline
M^{}_{\rm u}$ and $\overline M^{}_{\rm d}$ turns out to be
\begin{equation}\label{}
\overline M^{}_{\rm u}\simeq
 A^{}_{\rm u}\left(
 \begin{array}{ccc}
0 &\hspace{0.2cm}0.0005 &\hspace{0.2cm}0\\
 0.0005 &\hspace{0.2cm}1&\hspace{0.2cm}0.993\\
0 &\hspace{0.2cm}0.993 &\hspace{0.2cm}1
 \end{array}
 \right) \; ,
 \hspace{1cm}
 \overline M^{}_{\rm d}\simeq
 A^{}_{\rm d}\left(
 \begin{array}{ccc}
0 &\hspace{0.2cm}0.013 &\hspace{0.2cm} 0\\
 0.013 &\hspace{0.2cm}1 &\hspace{0.2cm}0.96\\
 0&\hspace{0.2cm}0.96 &\hspace{0.2cm}1
 \end{array}
 \right) \; .
\end{equation}
One can see that the (2,3) sectors of quark mass matrices are
suggestive of an underlying flavor symmetry which controls the
second and third quark families.

In fact, the $2\leftrightarrow 3$ permutation symmetry of quark mass
matrices, which is quite similar to the striking
$\mu\leftrightarrow\tau$ permutation symmetry in the lepton sector
\cite{XZ14}, has been conjectured long before \cite{fukuyama}. Under
this simple flavor symmetry the mass matrix takes the form
\begin{equation}\label{23symmetry} \overline M=
 \left(
 \begin{array}{ccc}
0 &\hspace{0.2cm}C &\hspace{0.2cm}C\\
 C &\hspace{0.2cm}A&\hspace{0.2cm}B\\
C &\hspace{0.2cm}B &\hspace{0.2cm}A
 \end{array}
 \right) \; .
\end{equation}
But such a scenario has been ruled out by the present experimental
data, as pointed out in Ref. \cite{23sym}. This situation can be
easily understood by taking a look at the expression of
$|V^{}_{ub}|$ in Eq.~(\ref{ckmelement}), where the two terms
originate from $O^{}_{\rm u}$ and $O^{}_{\rm d}$ in the following
way:
\begin{equation}\label{}
  \theta_{13}^{\rm d} \; \Longrightarrow \;
  \sqrt{\displaystyle \frac{m^{}_d}{m^{}_b}\
  \frac{m^{}_s}{m^{}_b}\left(\displaystyle \frac{1}{r^{}_{\rm d}}-1\right)}\ ,
   \hspace{1cm} \theta_{12}^{\rm u}\ V^{}_{cb} \; \Longrightarrow \;
   \sqrt{\displaystyle \frac{m^{}_u}{m^{}_c}}\ V^{}_{cb} \ .
\end{equation}
If there were an exact $2\leftrightarrow 3$ permutation symmetry, both
$\theta_{13}^{\rm u}$ and $\theta_{13}^{\rm d}$ would have to be vanishing.
However, the second term alone is unable to fit the experimental value
of $|V^{}_{ub}|$, as already discussed above.
Hence we conclude that quark mass matrices might possess a {\it partial}
$2\leftrightarrow 3$ permutation symmetry such that
\begin{equation}\label{}
  \overline{M}^{}_{22}=\overline{M}^{}_{33}\ ,\hspace{1cm}
  \overline{M}^{}_{12}\neq \overline{M}^{}_{13}\ .
\end{equation}
Since there is a large hierarchy between (1,2) and (3,3) entries of
$\overline{M}$ (i.e., $\overline{M}^{}_{33}\gg
\overline{M}^{}_{12}$), the $2\leftrightarrow 3$ permutation
symmetry can be taken as a starting point for model building, and it
is broken later on by introducing a small (1,2) entry. Furthermore,
the equality
$\overline{M}^{}_{22}=\overline{M}^{}_{23}=\overline{M}^{}_{33}$
should be a good leading-order approximation.

Having identified two special patterns of four-zero quark mass
matrices, we proceed to discuss the model building issues in order
to derive them. There are several ways to determine or constrain
quark flavor structures, among which flavor symmetries should be the
most popular and powerful one. So far a number of flavor symmetries,
such as the Abelian $U(1)$ flavor group \cite{fn} and the
non-Abelian $S(3)$ flavor group \cite{s3}, have been tried in this
respect. Before introducing a flavor symmetry to realize the above
special patterns of quark mass matrices, let us discuss
what the Hermiticity of $M^{}_{\rm u,d}$ implies for model building.

Quark mass matrices originate from the Yukawa interactions and are
in general non-Hermitian and complex. There are two possibilities of
making them Hermitian: (a) a proper transformation of the
right-handed quark fields, or equivalently a proper choice of the
flavor basis, as one has done in obtaining Eq. (2) or (3) in the SM
or its extensions which have no flavor-changing right-handed
currents; (b) imposing a reasonable assumption, such as the parity
symmetry to be discussed soon, on the Lagrangian of Yukawa
interactions. Note that case (a) is no more favored for our present
purpose, because an implementation of possible flavor symmetries is
also basis-dependent, and hence it is hard to coincide with the
chosen basis of Hermitian quark mass matrices in most cases. So let
us focus on case (b) in the following model-building exercises.

Under the parity symmetry, a flavor theory should be invariant
when a left-handed fermion field is replaced by its right-handed
counterpart (i.e., $\psi^{}_{\rm L} \to \psi^{}_{\rm R}$),
or vice versa. As for the Yukawa interactions of quark fields,
the parity transformation is
\begin{equation}\label{}
  y^{}_{ij} \overline{\psi_{\rm L}^i} \langle H\rangle \psi_{\rm R}^j
  + y_{ij}^*\overline{\psi_{\rm R}^j} \langle H\rangle \psi_{\rm L}^i
  \hspace{0.5cm} \longleftrightarrow \hspace{0.5cm}
  y^{}_{ij}\overline{\psi_{\rm R}^i} \langle H\rangle \psi_{\rm L}^j
  + y_{ij}^*\overline{\psi_{\rm L}^j} \langle H\rangle \psi_{\rm R}^i\ ,
\end{equation}
where $i$ and $j$ are the quark flavor indices, and $\langle
H\rangle$ stands for the vacuum expectation value (VEV) of the Higgs
field. The invariance of Yukawa interactions under parity
transformation requires the Yukawa coupling matrix elements to
satisfy the condition $y^{}_{ij}=y_{ji}^*$, and hence the
corresponding quark mass matrix must be Hermitian in the flavor
space. We are therefore motivated to consider Hermitian quark mass
matrices in the framework of the Left-Right (LR) symmetric model
with an explicit parity symmetry \cite{lr}.

The LR model extends the SM gauge groups to $SU(2)^{}_{\rm L} \times
SU(2)^{}_{\rm R} \times U(1)^{}_{B-L}$, where $SU(2)^{}_{\rm R}$ is
the opposite of $SU(2)^{}_{\rm L}$ and acts only on the iso-doublets
constituted by the right-handed fields, and $B$$-$$L$ stands for the
baryon number minus the lepton number. All the fermion fields are
grouped into iso-doublets as follows:
\begin{equation}\label{}
Q_{\rm L}^i =\left(\begin{array}{c}
    u_{\rm L}^i \\ d_{\rm L}^i
 \end{array}
 \right) \; , \hspace{1cm}
 Q_{\rm R}^i =\left(\begin{array}{c}
    u_{\rm R}^i\\ d_{\rm R}^i
 \end{array}
 \right) \; ,\hspace{1cm}
 L_{\rm L}^i =\left(\begin{array}{c}
    \nu_{\rm L}^i \\ e_{\rm L}^i
 \end{array}
 \right) \; , \hspace{1cm}
  L_{\rm R}^i =\left(\begin{array}{c}
    \nu_{\rm R}^i \\ e_{\rm R}^i
 \end{array}
 \right) \; .
\end{equation}
In the present work we concentrate on the quark sector and leave out
the lepton fields $L_{\rm L}^i$ and $L_{\rm R}^i$. At the scale
$\Lambda^{}_{\rm R}$ which is higher than the electroweak scale,
$SU(2)^{}_{\rm R} \times U(1)^{}_{B-L}$ is broken to $U(1)^{}_{\rm
Y}$. The residual $SU(2)^{}_{\rm L}$ and $U(1)^{}_{\rm Y}$ are
exactly the SM gauge groups which are subsequently broken by a
bi-doublet field $\Phi$ under $SU(2)^{}_{\rm L} \times SU(2)^{}_{\rm
R}$:
\begin{equation}\label{}
  \Phi=\left(\begin{array}{cc}
    \phi_1^0& ~\phi_2^{+}\\
     \phi_1^{-}& ~\phi_2^0
 \end{array}\right)
 \hspace{0.45cm} \longrightarrow {\rm VEV} \longrightarrow
 \hspace{0.45cm}
 \langle\Phi\rangle=\left(\begin{array}{cc}
    \kappa& \\
     & \kappa^\prime
 \end{array}\right) \; .
\end{equation}
The six quarks acquire their masses via their Yukawa interactions
with $\Phi$:
\begin{equation}\label{}
  \overline{\left(u_{\rm L}^i ~~ d_{\rm L}^i\right)}
  \left[y^{}_{ij}\left(\begin{array}{cc}
    \kappa& \\
     & \kappa^\prime
 \end{array}\right)+y_{ij}^\prime\left(\begin{array}{cc}
    \kappa^\prime& \\
     & \kappa
 \end{array}\right) \right]\left(\begin{array}{c}
    u_{\rm R}^j\\ d_{\rm R}^j
 \end{array}
 \right) + {\rm h.c.} \ .
\end{equation}
In the {\it minimal} non-supersymmetric LR model $\kappa^\prime$ has a relative
phase as compared with $\kappa$, and this may violate the
Hermiticity of quark mass matrices. Hence we prefer to (but not necessarily)
work in the framework of the supersymmetric (SUSY) LR model \cite{susylr}. Note
that the $y_{ij}^\prime$ term in Eq. (35) will be forbidden by the
holography requirement of the superpotential in this framework.
\begin{table}
\caption{The fields relevant for the Yukawa couplings and their charges
under $U(1)^{}_{\rm FN}$.}
\begin{tabular}
{|p{2cm}<{\centering}|p{2cm}<{\centering}|p{2cm}<{\centering}|
p{1.5cm}<{\centering}|
p{1.5cm}<{\centering}|p{1.5cm}<{\centering}|p{1.5cm}<{\centering}|
p{1.5cm}<{\centering}|}
\hline
$Q_{\rm L}^1/Q_{\rm R}^{1c}$ &$Q_{\rm L}^2/Q_{\rm R}^{2c}$ &
$Q_{\rm L}^3/Q_{\rm R}^{3c}$ & $\Phi^{}_1$  & $\Phi^{}_2$ &
$\Phi^{}_3$/$\Phi^{}_4$ & $S^{}_1$ & $S^{}_2$ \\
         \hline
$-5$ & $4$ & $0$ & $0$ & $-1$ & $1$ & $-4$ & $-1$ \\
         \hline
\end{tabular}
\label{u1}
\end{table}

Now that the issue of Hermiticity has been settled, let us continue
to build quark mass models under certain flavor symmetries in a
usual way. We begin with a model that can lead to a four-zero
texture of $M^{}_{\rm u}$ and $M^{}_{\rm d}$ in the $r\sim 1$
region. It is easy to derive the special pattern of $M^{}_{\rm u,d}$
in Eq.~(\ref{2309}) with the help of the Froggatt-Nielsen (FN)
mechanism \cite{fn}. The point is to introduce a global
$U(1)^{}_{\rm FN}$ symmetry to structure the quark mass matrices.
All the fields relevant for quark masses and their charges under
$U(1)^{}_{\rm FN}$ are listed in TABLE \ref{u1}. According to the
convention in SUSY, $Q_{\rm R}^i$ is represented by its
corresponding left-handed chiral superfield $Q_{\rm R}^{ic}$. In the
SUSY LR models, at least two bi-doublets are needed to avoid the
exact parallelism between $M^{}_{\rm u}$ and $M^{}_{\rm d}$. In our
model four bi-doublets are introduced, and their VEVs are written
as
\begin{eqnarray}\label{vevs}
&& \langle\Phi^{}_1\rangle=\left(\begin{array}{cc}
    \kappa^{}_1& \\
     & \kappa_1^\prime
 \end{array}\right) \; , \hspace{1cm}
  \langle\Phi^{}_2\rangle=\left(\begin{array}{cc}
    \kappa^{}_2& \\
     & \kappa_2^\prime
 \end{array}\right) \; , \nonumber \\
&& \langle\Phi^{}_3\rangle=\left(\begin{array}{cc}
    \kappa^{}_3& \\
     & \kappa_3^\prime
 \end{array}\right) \; , \hspace{1cm}
  \langle\Phi^{}_4\rangle=\left(\begin{array}{cc}
    \kappa^{}_4& \\
     & \kappa_4^\prime
 \end{array}\right) \; .
\end{eqnarray}
In addition, two gauge singlets $S^{}_1$ and $S^{}_2$ are introduced
to spontaneously break the $U(1)^{}_{\rm FN}$ flavor symmetry.

For clarity, let us explore the phenomenological consequences of this model
step by step. The contribution from $\Phi^{}_{1}$ can be expressed as
\begin{equation}\label{}
\begin{array}{l}
\vspace{0.3cm}
  y^{}_{33}Q_{\rm R}^{3c}\Phi^{}_{1}Q_{\rm L}^{3} + y^{}_{23}
  Q_{\rm R}^{2c}\Phi^{}_{1}Q_{\rm L}^{3}\displaystyle \frac{S^{}_1}{\Lambda}
  + y_{23}^*Q_{\rm R}^{3c}\Phi^{}_{1}Q_{\rm L}^{2}
  \frac{S^{}_1}{\Lambda} + y^{}_{22}Q_{\rm R}^{2c}\Phi^{}_{1}Q_{\rm L}^{2}
  \displaystyle \left(\frac{S^{}_1}{\Lambda}\right)^2 \ ,
  \end{array}
\end{equation}
where $y^{}_{22}$ and $y^{}_{33}$ are real, but $y^{}_{23}$ is
complex. $\Lambda$ is the scale where all the fields associated with
the FN mechanism reside. The non-renormalizable operators arise from
integrating out the heavy fields which are not explicitly given in
TABLE \ref{u1}, and thus they are suppressed by $\Lambda$. The key
point of the FN mechanism is to assume that the ratios of $\langle
S^{}_1\rangle$ and $\langle S^{}_2\rangle $ to $\Lambda$ are small
quantities which can be generally denoted as $\epsilon$, such that
each element of quark mass matrices is encoded in a power of
$\epsilon$. Here we have identified this small quantity with the one
in Eq. (\ref{2309}), and thus its magnitude is about 0.3. When
$S^{}_1$ and $\Phi^{}_1$ acquire their VEVs, the (2,3) sectors of
$M^{}_{\rm u}$ and $M^{}_{\rm d}$ are of the form
\begin{equation}\label{lo}
M_{\rm u}^{(2,3)}\sim
 y^{}_{33}\kappa^{}_1\left(
 \begin{array}{cc}
 \vspace{0.3cm}
\displaystyle \frac{y^{}_{22}}{y^{}_{33}}\epsilon^2
&\hspace{0.3cm}\displaystyle \frac{y^{}_{23}}{y^{}_{33}}\epsilon \\
\displaystyle \frac{y_{23}^*}{y^{}_{33}}\epsilon &\hspace{0.3cm}1
 \end{array}
 \right) \; ,
 \hspace{1cm}
 M_{\rm d}^{(2,3)}\sim
 y^{}_{33}\kappa_1^\prime\left(
 \begin{array}{cc}
  \vspace{0.3cm}
\displaystyle \frac{y^{}_{22}}{y^{}_{33}}\epsilon^2
&\hspace{0.3cm}\displaystyle \frac{y^{}_{23}}{y^{}_{33}}\epsilon \\
\displaystyle \frac{y_{23}^*}{y^{}_{33}}\epsilon &\hspace{0.3cm}1
 \end{array}
 \right) \; ,
\end{equation}
which can reproduce the flavor structure in Eq.~(\ref{2309}). There
is the exact parallelism between up and down quark sectors, because
they have the same origin (i.e., from $\Phi^{}_1$ here). However,
this situation also brings about two phenomenological problems. One
of them is that the (2,2) entries of $M^{}_{\rm u}$ and $M^{}_{\rm
d}$ actually do not respect this exact parallelism, as we have seen
in Eq.~(\ref{r09}). The other problem is that the (2,3) entries of
$M^{}_{\rm u}$ and $M^{}_{\rm d}$ should have a phase difference, so
as to assure $\phi^{}_2 \neq 0$ or $2\pi$.

To address these two problems, let us take
account of the contribution from $\Phi^{}_{2}$ as follows:
\begin{equation}\label{}
\begin{array}{l}
\vspace{0.3cm}
  y_{23}^\prime Q_{\rm R}^{2c}\Phi^{}_{2} Q_{\rm L}^{3}\displaystyle
  \left(\frac{S^{}_2}{\Lambda}\right)^3
  +y_{23}^{\prime *}Q_{\rm R}^{3c}\Phi^{}_{2}Q_{\rm L}^{2} \displaystyle
  \left(\frac{S^{}_2}{\Lambda}\right)^3
  +y_{22}^\prime Q_{\rm R}^{2c}\Phi^{}_{2}Q_{\rm L}^{2}\displaystyle
  \frac{S^{}_1 S_2^3}{\Lambda^4} \ .
  \end{array}
\end{equation}
This treatment modifies Eq.~(\ref{lo}) to the form
\begin{equation}\label{}
\begin{array}{l}
\vspace{0.5cm}
M_{\rm u}^{(2,3)}\sim
 y^{}_{33}\kappa^{}_1\left(
 \begin{array}{cc}
 \vspace{0.3cm}
\displaystyle \frac{y^{}_{22}}{y^{}_{33}}\epsilon^2+\displaystyle
\frac{y_{22}^\prime}{y^{}_{33}}\displaystyle \frac{\kappa^{}_2}
{\kappa^{}_1}\epsilon^4
&\hspace{0.3cm}\displaystyle \frac{y^{}_{23}}{y^{}_{33}}\epsilon
+ \displaystyle \frac{y_{23}^\prime}{y^{}_{33}}\displaystyle
\frac{\kappa^{}_2}{\kappa^{}_1}\epsilon^3\\
\displaystyle \frac{y_{23}^*}{y^{}_{33}}\epsilon
+\displaystyle \frac{y_{23}^{\prime *}}{y^{}_{33}}\displaystyle
\frac{\kappa^{}_2}{\kappa^{}_1}\epsilon^3
&\hspace{0.3cm}1
 \end{array}
 \right) \; , \vspace{-0.2cm} \\
M_{\rm d}^{(2,3)}\sim
 y^{}_{33}\kappa_1^\prime\left(
 \begin{array}{cc}
 \vspace{0.3cm}
\displaystyle \frac{y^{}_{22}}{y^{}_{33}}\epsilon^2
+\displaystyle \frac{y_{22}^\prime}{y^{}_{33}}\displaystyle
\frac{\kappa_2^\prime}{\kappa_1^\prime}\epsilon^4
&\hspace{0.3cm}\displaystyle \frac{y^{}_{23}}{y^{}_{33}}\epsilon
+ \displaystyle \frac{y_{23}^\prime}{y^{}_{33}}\displaystyle
\frac{\kappa_2^\prime}{\kappa_1^\prime}\epsilon^3\\
\displaystyle \frac{y_{23}^*}{y^{}_{33}}\epsilon
+\displaystyle \frac{y_{23}^{\prime *}}{y^{}_{33}}\displaystyle
\frac{\kappa_2^\prime}{\kappa_1^\prime}\epsilon^3
&\hspace{0.3cm}1
 \end{array}
 \right) \; .
 \end{array}
\end{equation}
If the ratios $\kappa^{}_1/\kappa^{}_2$ and
$\kappa_1^\prime/\kappa_2^\prime$ are close but not exactly equal to
each other, the difference between the (2,2) entries of $M^{}_{\rm
u}$ and $M^{}_{\rm d}$ will be of ${\cal O}(\epsilon^4) \sim 0.01$,
in agreement with the numerical result given in Eq.~(\ref{r09}). The
difference between the (2,3) entries of $M^{}_{\rm u}$ and $M^{}_{\rm
d}$ seems to be of ${\cal O}(\epsilon^3) \sim 0.03$ and in conflict
with Eq.~(\ref{r09}). One may essentially get around this problem by
assuming that the phase difference between $y^{}_{23}$ and
$y^\prime_{23}$ is about $\pi/2$, such that the absolute values of
$M^{}_{\rm u 23}$ and $M^{}_{\rm d 23}$ only have a negligibly small
difference of ${\cal O}(\epsilon^5) \sim 0.003$. But the phase
difference between $M^{}_{\rm u 23}$ and $M^{}_{\rm d 23}$ is of
${\cal O}(\epsilon^2) \sim 0.1$, just consistent with the value of
$\phi^{}_2$ illustrated in section \Rmnum 2.

Finally, $\Phi^{}_{3}$ and $\Phi^{}_{4}$ can offer finite masses for
the first quark family through the terms
\begin{equation}\label{}
\begin{array}{l}
\vspace{0.3cm}
  y^{}_{12}Q_{\rm R}^{1c}\Phi^{}_{3}Q_{\rm L}^{2}
  + y_{12}^*Q_{\rm R}^{2c}\Phi^{}_{3}Q_{\rm L}^{1}
  +y_{12}^\prime Q_{\rm R}^{1c}\Phi^{}_{4} Q_{\rm L}^{2}
  +y_{12}^{\prime *}Q_{\rm R}^{2c}\Phi^{}_{4} Q_{\rm L}^{1}\ ,
  \end{array}
\end{equation}
from which we obtain
\begin{equation}\label{}
 M^{}_{\rm u 12}=M_{\rm u 21}^* = y^{}_{12}\kappa^{}_3
 + y_{12}^\prime \kappa^{}_4 \; , \hspace{1cm}
 M^{}_{\rm d 12}=M_{\rm d 21}^* = y^{}_{12} \kappa_{3}^\prime
 +y_{12}^\prime \kappa_{4}^\prime\ .
\end{equation}
The reason that we arrange $\Phi^{}_3$ and $\Phi^{}_4$ to have the
same quantum number is rather simple: in this case the phases of
$M^{}_{\rm u 12}$ and $M^{}_{\rm d 12}$ can be different, such that
we are left with a nonzero $\phi^{}_1$. In a complete
flavor-symmetry model the smallness of $M^{}_{\rm u 12}$ and
$M^{}_{\rm d 12}$ should also be explained via the FN mechanism as
we have done for the (2,3) sectors of $M^{}_{\rm u}$ and $M^{}_{\rm
d}$. Instead of repeating a similar exercise, here we simply assume
that $\kappa^{}_3$, $\kappa_3^\prime$, $\kappa^{}_4$ and
$\kappa_4^\prime$ are much smaller than their counterparts
$\kappa^{}_1$, $\kappa_1^\prime$, $\kappa^{}_2$ and
$\kappa_2^\prime$. Of course, the elements $M^{}_{11}$ and
$M^{}_{13}$ are vanishing as limited by the relevant flavor quantum
numbers.

When it comes to the particular case $M^{}_{22}=M^{}_{33}$, a
non-Abelian flavor symmetry is needed to realize this equality. The
simplest candidate of this kind is the $S(3)$ group which has three
irreducible representations ${\textbf{1}}$, ${\textbf{1$^\prime$}}$
and ${\textbf{2}}$. The tensor products of these representations can
be decomposed as follows \cite{s3rule}:
\begin{equation}\label{}
  \begin{array}{l}
  \vspace{0.5cm}
   \left(\begin{array}{c}
    x^{}_1\\x^{}_2
  \end{array}\right)_{\textbf 2}  \times
   \left(\begin{array}{c}
    y^{}_1\\y^{}_2
  \end{array}\right)_{\textbf 2}=
  \left(x^{}_1 y^{}_1 + x^{}_2 y^{}_2\right)_{\textbf 1}
  + \left(x^{}_1 y^{}_2 - x^{}_2 y^{}_1\right)_{\textbf {1$^\prime$}}+
   \left(\begin{array}{c}
    x^{}_1 y^{}_2 + x^{}_2 y^{}_1 \\ x^{}_1 y^{}_1 - x^{}_2 y^{}_2
  \end{array}\right)_{\textbf 2} \; ,\\
  \left(\begin{array}{c}
    x^{}_1 \\ x^{}_2
  \end{array}\right)_{\textbf 2}\times y^{}_{\textbf {1$^\prime$}} =
   \left(\begin{array}{c}
    -x^{}_2 y \\x^{}_1 y
  \end{array}\right)_{\textbf 2} \; , \hspace{1cm}
   x^{}_{\textbf {1$^\prime$}}\times
   y^{}_{\textbf {1$^\prime$}}=\left(xy\right)^{}_{\textbf 1} \ .
  \end{array}
\end{equation}

The quark fields are organized to be the representations of $S(3)$
in the following way:
\begin{equation}\label{}
   Q_{\rm L}^{1}-{\bf 1}\ , \hspace{0.5cm} Q_{\rm R}^{1c}-{\bf 1}\
   ,\hspace{0.5cm}
  \left(\begin{array}{c}
    Q_{\rm L}^{2} \\ Q_{\rm L}^{3}
  \end{array}\right)-{\bf 2}\ ,\hspace{0.5cm}
  \left(\begin{array}{c}
    Q_{\rm R}^{2c} \\ Q_{\rm R}^{3c}
  \end{array}\right)-{\bf 2}\ ,
\end{equation}
while the bi-doublets introduced and their representations under
the $S(3)$ group are:
\begin{equation}\label{}
  \Phi^{}_1-{\bf 1}\ ,\hspace{0.5cm}
   \Phi^{}_2-{\bf 1^{\prime}}\ ,\hspace{0.5cm}
  \left(\begin{array}{c}
    \Phi^{}_3 \\ \Phi^{}_{4}
  \end{array}\right)-{\bf 2}\ .
\end{equation}
The VEVs of bi-doublets are specified to be
\begin{eqnarray}\label{}
&& \langle\Phi^{}_1\rangle=\left(\begin{array}{cc}
    \kappa^{}_1& \\
     & \kappa_1^\prime
 \end{array}\right) \; , \hspace{1cm}
  \langle\Phi^{}_2\rangle=\left(\begin{array}{cc}
    \kappa^{}_2& \\
     & \kappa_2^\prime
 \end{array}\right) \; , \nonumber \\
&& \langle\Phi^{}_3\rangle=\left(\begin{array}{cc}
    \kappa^{}_3& \\
     & \kappa_3^\prime
 \end{array}\right) \; , \hspace{1cm}
  \langle\Phi^{}_4\rangle=\left(\begin{array}{cc}
    0&\hspace{0.4cm} \\
     &\hspace{0.4cm} 0
 \end{array}\right) \; .
\end{eqnarray}
In this model the equality of $M^{}_{22}$ and $M^{}_{33}$ results from
the operator
\begin{equation}\label{2233}
  y^{}_1 \left(\begin{array}{c}
    Q_{\rm R}^{2c}\\Q_{\rm R}^{3c}
  \end{array}\right)\Phi^{}_1
  \left(\begin{array}{c}
    Q_{\rm L}^{2}\\Q_{\rm L}^{3}
  \end{array}\right)
~  \Longrightarrow ~ y^{}_1 \left[ Q_{\rm R}^{2c}
\langle\Phi^{}_1 \rangle Q_{\rm L}^{2}+Q_{\rm R}^{3c}
\langle\Phi^{}_1\rangle  Q_{\rm L}^{3} \right] \; ,
\end{equation}
and their values are given by
\begin{equation}\label{}
  M^{}_{\rm u 22}= M^{}_{\rm u 33} = y^{}_1 \kappa^{}_1 \ ,
  \hspace{1cm} M^{}_{\rm d 22} = M^{}_{\rm d 33} = y^{}_1 \kappa_1^\prime\
  .
\end{equation}
In comparison, the elements $M^{}_{23}$ and $M^{}_{32}$ are
generated by the operators
\begin{equation}\label{2332}
\begin{array}{ll}
\vspace{0.2cm}
  y^{}_2 \left(\begin{array}{c}
    Q_{\rm R}^{2c}\\Q_{\rm R}^{3c}
  \end{array}\right)
  \Phi^{}_2
  \left(\begin{array}{c}
    Q_{\rm L}^{2}\\Q_{\rm L}^{3}
  \end{array}\right)
&~ \Longrightarrow ~ y^{}_2\left[Q_{\rm R}^{2c}\langle \Phi^{}_2
\rangle Q_{\rm L}^{3}+Q_{\rm R}^{3c}\langle \Phi^{}_2 \rangle Q_{\rm
L}^{2}\right] \; ,\\
  y^{}_3 \left(\begin{array}{c}
    Q_{\rm R}^{2c}\\Q_{\rm R}^{3c}
  \end{array}\right)
  \left(\begin{array}{c}
    \Phi^{}_3\\ \Phi^{}_{4}
  \end{array}\right)
  \left(\begin{array}{c}
    Q_{\rm L}^{2}\\Q_{\rm L}^{3}
  \end{array}\right)
&~ \Longrightarrow ~ y^{}_3\left[Q_{\rm R}^{2c}\langle \Phi^{}_3
\rangle Q_{\rm L}^{3}+Q_{\rm R}^{3c}\langle \Phi^{}_4 \rangle Q_{\rm
L}^{2}\right] \; ,
\end{array}
\end{equation}
which lead us to
\begin{equation}\label{}
\begin{array}{l}
  M^{}_{\rm u 23}= y^{}_2 \kappa^{}_2+y^{}_3 \kappa^{}_3 \hspace{1cm}
  M^{}_{\rm u 32}= -y^{}_2 \kappa^{}_2+y^{}_3 \kappa^{}_3\ ,\\
  M^{}_{\rm d 23}= y^{}_2 \kappa_2^\prime+y^{}_3 \kappa^{\prime}_3 \hspace{1cm}
  M^{}_{\rm d 32} = -y^{}_2 \kappa_2^\prime +y^{}_3 \kappa^{\prime}_3\ .
  \end{array}
\end{equation}
Notice that the Hermiticity of quark mass matrices as required by the
LR symmetry leaves us $y_2^*=-y_2^{}$ and $y_3^*=y_3^{}$ (i.e., $y_2^{}$
is imaginary while $y_3^{}$ is real). If only one of the $y_2^{}$ and
$y_3^{}$ terms exists, $\phi_2$ will be zero, so both of them are necessary.
Another noteworthy point is that the operators in
Eqs.~(\ref{2233}) and (\ref{2332}) are completely independent of
each other, so it is difficult to understand why $M^{}_{23}$ is so
close to $M^{}_{22}$ and $M^{}_{33}$. We conjecture that these two
operators are possible to come from the same tensor product in a
larger group, so that $M^{}_{22}=M^{}_{33}=M^{}_{23}$ can be
obtained as the leading-order approximation.

Finally, let us consider the operators
\begin{equation}\label{}
\begin{array}{l}
  y^{}_3  Q_{\rm R}^{1c}
  \left(\begin{array}{c}
    \Phi^{}_2\\ \Phi^{}_{3}
  \end{array}\right)
  \left(\begin{array}{c}
    Q_{\rm L}^{2}\\Q_{\rm L}^{3}
  \end{array}\right)
  + y_3^*   \left(\begin{array}{c}
    Q_{\rm R}^{2c}\\Q_{\rm R}^{3c}
  \end{array}\right)
  \left(\begin{array}{c}
    \Phi^{}_2\\ \Phi^{}_{3}
  \end{array}\right)
Q_{\rm L}^{1}
\hspace{0.05cm} \Longrightarrow \hspace{0.05cm}
 y^{}_3  Q_{\rm R}^{1c}\langle \Phi^{}_2 \rangle Q_{\rm L}^2
 + y_3^* Q_{\rm R}^{2c}\langle \Phi^{}_2 \rangle Q_{\rm L}^1\ .
\end{array}
\end{equation}
They lead us to the nonzero (1,2) entries of quark mass matrices:
\begin{equation}\label{1221}
  M^{}_{\rm u 12}= M_{\rm u 21}^*=y^{}_3 \kappa^{}_2\ ,
  \hspace{1cm} M^{}_{\rm d 12}= M_{\rm d 21}^*=y^{}_3 \kappa_2^\prime\ .
\end{equation}
Note that it is $\langle \Phi^{}_3 \rangle =0$ that ensures the
vanishing of $M^{}_{13}$ and $M^{}_{31}$.
There is also a problem that $\phi^{}_1$ equals zero, but it can be
overcome by introducing the column vector $(\Phi^{}_4 ,
\Phi^{}_5)^{\rm T}$. Similar to $(\Phi^{}_2 , \Phi^{}_3)^{\rm T}$,
$\Phi^{}_4$ acquires its VEV but $\Phi^{}_5$ does not. In this case
Eq.~(\ref{1221}) is modified to the form
\begin{equation}\label{1221m}
  M^{}_{\rm u 12}= M_{\rm u 21}^*=y^{}_3 \kappa_2 + y_3^\prime
  \kappa^{}_4\ ,\hspace{1cm}
  M^{}_{\rm d 12}= M_{\rm d 21}^*=y^{}_3 \kappa_2^\prime +
  y_3^\prime \kappa_4^\prime\ .
\end{equation}
We just need $\kappa^{}_4/\kappa^{}_2 \neq
\kappa_4^\prime/\kappa_2^\prime$ to make $\phi^{}_1$ nonzero. The
last remarkable issue is that one needs to impose the FN quantum
numbers on $Q_{\rm L}^1$ and $Q_{\rm R}^{1c}$, in order to explain
why the magnitude of $M^{}_{12}$ is suppressed by a power of
$\epsilon$ as compared with those of $M^{}_{22}$ and $M^{}_{23}$.
Such a treatment can also help avoid a large $M^{}_{11}$ arising
from the operator $y^{}_{5}Q_{\rm R}^{1c}\langle \Phi^{}_{1}\rangle
Q_{\rm L}^1$. If we assign an FN quantum number $n$ to both $Q_{\rm
L}^1$ and $Q_{\rm R}^{1c}$, for instance, the contribution of this
operator will be suppressed by $\epsilon^{2n}$ and thus negligibly
small.

In short, we have identified two special four-zero patterns of quark
mass matrices and discussed two toy models for realizing them. We
should point out that the introduction of so many bi-doublet Higgs fields
may cause the flavor-changing-neutral-current (FCNC) problem.
But this problem can be avoided by assuming that the LR symmetry
breaks at a very high scale and there is just one (two) effective
Higgs field(s) (as linear combinations of the above Higgs fields) at the low
scale in which case we go back to the SM (MSSM) situation. Otherwise,
we can address this issue by introducing some flavon fields located
at a superhigh energy scale to play the role of bi-doublets as multiple
representations of the flavor symmetries. In this case, we do not need Higgs
fields other than the usual ones which have already been required for other
purposes rather than the flavor physics. After integrating out the flavon
fields, there will be no trace of the flavor physics except that the Yukawa
couplings have been constrained by the flavor symmetries. This way of preventing
the flavor physics from disturbing the other physics is widely used
in flavor-symmetry models for the lepton sector \cite{f-symmetry}.

\section{On the stability of the four-zero texture}

As shown in section III, the four-zero texture of quark mass
matrices may result from an underlying flavor symmetry. But the
failure in discovering any new physics of this kind indicates that
it is likely to reside in a superhigh energy scale, such as the
grand unification theory (GUT) scale. This means that a
flavor-symmetry model should be built somewhere far above the
electroweak scale and the RGE running effects have to be taken into
account when its phenomenological consequences are confronted with
the experimental data at low energies \cite{renorm}. One may follow
two equivalent ways to consider the evolution of energy scales,
provided there is no new physics between the flavor symmetry scale
$\Lambda^{}_{\rm FS}$ and the electroweak scale $M^{}_Z$
\cite{lindner}: (a) the first step is to figure out quark masses and
flavor mixing parameters from $M^{}_{\rm u}$ and $M^{}_{\rm d}$ at
$\Lambda^{}_{\rm FS}$, and the second step is to run these physical
quantities down to $M^{}_Z$ via their RGEs; (b) the first step is to
evolve $M^{}_{\rm u}$ and $M^{}_{\rm d}$ from $\Lambda^{}_{\rm FS}$
down to $M^{}_Z$ via their RGEs, and the second step is to calculate
quark masses and flavor mixing parameters from the corresponding
quark mass matrices at $M^{}_Z$. Here we take advantage of way (b)
to examine the stability of texture zeros of $M^{}_{\rm u}$ and
$M^{}_{\rm d}$ against the evolution of energy scales in an
analytical way. The RGE effect on the Fritzsch texture of quark mass
matrices has been studied in a similar way \cite{lindner,lindner0}.

\begin{figure}
\includegraphics[width=4.35in]{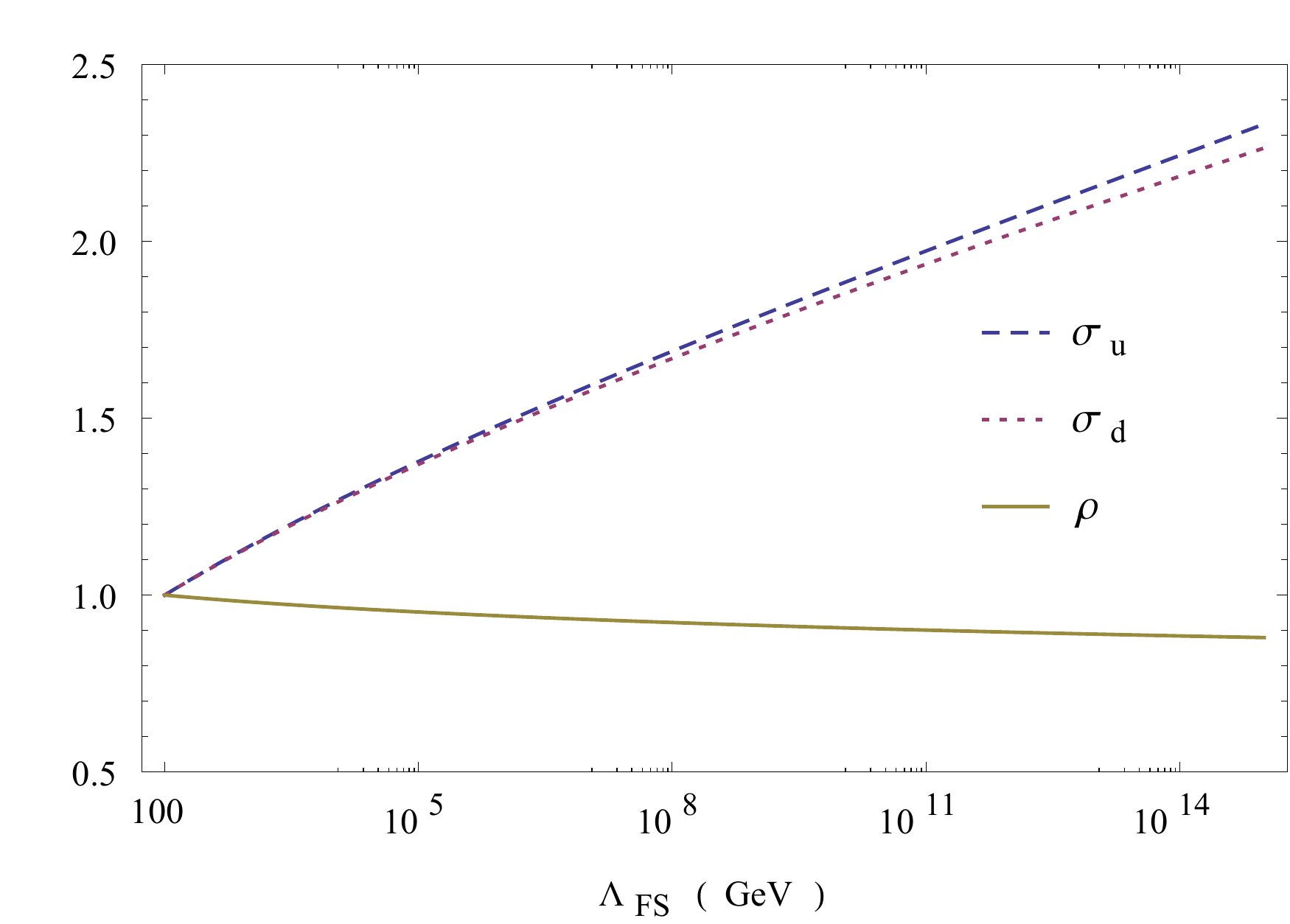}
\caption{An illustration of the changes of $\sigma^{}_{\rm q}$ and $\rho$
with the scale $\Lambda^{}_{\rm FS}$ in the SM.} \label{sigmarho}
\end{figure}

At the one-loop level, the RGEs of the quark Yukawa coupling matrices
in the SM can be written as
\begin{equation}\label{rge}
  16\pi^2 \displaystyle \frac{{\rm d} Y^{}_{\rm q}(t)}{{\rm d} t}
  =\left[\ \displaystyle \frac{3}{2}S^{}_{\rm q}(t)-G^{}_{\rm q}(t)\bm{1}
  + T(t)\bm{1}\ \right]Y^{}_{\rm q}(t)\ ,
\end{equation}
where $t=\ln(\mu/M^{}_{Z})$, and the subscript ``q" stands for ``u"
and ``d". The contributions of the charged leptons and neutrinos to
Eq.~(\ref{rge}) have been omitted, because they are negligibly small
in the SM. Denoting the VEV of the Higgs field as $v$, we can
express the four-zero texture of $Y^{}_{\rm u}$ and $Y^{}_{\rm d}$
at $\Lambda^{}_{\rm FS}$ as follows:
\begin{equation}\label{repara}
\begin{array}{l}
Y^{}_{\rm u}(\Lambda^{}_{\rm {FS}})=\displaystyle
\frac{1}{v}M^{}_{{\rm u}}(\Lambda^{}_{\rm {FS}})=\left(
 \begin{array}{ccc}
 0&\hspace{0.1cm} c^{}_{\rm u} &\hspace{0.1cm} 0\\
c^{}_{\rm u} &\hspace{0.1cm} \tilde{b}^{}_{\rm u} & \hspace{0.1cm}
b^{}_{\rm u}\\
0&\hspace{0.1cm} b^{}_{\rm u} &\hspace{0.1cm} a^{}_{\rm u}
 \end{array}
 \right) , \hspace{0.3cm}
 Y^{}_{\rm d}(\Lambda^{}_{\rm {FS}})=\displaystyle
 \frac{1}{v} M^{}_{{\rm d}}(\Lambda^{}_{\rm {FS}})=\left(
 \begin{array}{ccc}
 0&\hspace{0.1cm} c^{}_{\rm d} &\hspace{0.1cm} 0\\
c_{\rm d}^* &\hspace{0.1cm} \tilde{b}_{\rm d} &
\hspace{0.1cm} b^{}_{\rm d}\\
0&\hspace{0.1cm} b_{\rm d}^* &\hspace{0.1cm} a^{}_{\rm d}
 \end{array}
 \right) .
 \end{array}
\end{equation}
Without loss of generality for CP violation, we have chosen
$b^{}_{\rm u}$ and $c^{}_{\rm u}$ to be real in Eq.~(\ref{repara}).
The terms $G^{}_{\rm q}(t)$ and $T(t)$ read
\begin{equation}\label{}
  G^{}_{\rm u}=G^{}_{\rm d}+g_1^2= 8g_3^2+\displaystyle
  \frac{9}{4}g_2^2+\displaystyle \frac{17}{12}g_1^2\ ,
   \hspace{1cm} T=3 {\rm {Tr}}\left(Y^{}_{\rm u}Y_{\rm u}^\dagger
   +Y^{}_{\rm d}Y_{\rm d}^\dagger\right)\ ,
\end{equation}
which arise from quantum corrections to the quark and Higgs field
strengths, respectively. They are flavor-blind, and thus
proportional to the identity matrix in the flavor space. Since their
effects are simply to rescale quark mass matrices as a whole at a
lower energy scale, they will be dropped for the moment. Namely, we
are mainly concerned about the first term in Eq.~(\ref{rge}):
$S^{}_{\rm u}=-S^{}_{\rm d}=Y^{}_{\rm u}Y_{\rm u}^\dagger-Y^{}_{\rm
d}Y_{\rm d}^\dagger$, which governs the nonlinear evolution of
$Y^{}_{\rm q}$. Defining $H^{}_{\rm q} = Y^{}_{\rm q}Y_{\rm
q}^\dagger$, let us rewrite Eq.~(\ref{rge}) by dropping its
$G^{}_{\rm q}(t)$ and $T(t)$ terms:
\begin{equation}\label{rge2}
  \displaystyle \frac{32\pi^2}{3}\displaystyle
  \frac{{\rm d} H^{}_{\rm q}}{{\rm d} t} = S^{}_{\rm q}H^{}_{\rm q}
  + H^{}_{\rm q}S^{}_{\rm q}\ .
\end{equation}
In a good approximation $S^{}_{\rm q}$ can be expressed as
 \begin{equation}\label{}
 S^{}_{\rm u}=-S^{}_{\rm d}\simeq \left(
 \begin{array}{ccc}
 0&\hspace{0.3cm} 0 &\hspace{0.3cm} 0 \\
0 &\hspace{0.3cm} \Delta^{}_2 &
\hspace{0.3cm} \Delta^{}_3 \\
0&\hspace{0.3cm} \Delta^{}_3
&\hspace{0.3cm} \Delta^{}_1
 \end{array}
 \right) \; ,
\end{equation}
where $\Delta^{}_1=a_{\rm u}^2+b_{\rm u}^2$,
$\Delta^{}_2=b_{\rm u}^2+\tilde b_{\rm u}^2$ and
$\Delta^{}_3=b^{}_{\rm u} (a^{}_{\rm u}+\tilde{b}^{}_{\rm u})$.
Then we solve the differential equations in Eq.~(\ref{rge2}) and
obtain
\footnotesize
\begin{equation}\label{}
H^{}_{\rm u}(M^{}_Z) \simeq \left(
 \begin{array}{ccc}
 \vspace{0.3cm}
 c_{\rm u}^2 &\hspace{0.2cm}c^{}_{\rm u}\tilde{b}^{}_{\rm u} \rho
 -\displaystyle \frac{\rho -1}{\Delta^{}_1+\Delta^{}_2}
 c^{}_{\rm u} a^{}_{\rm u}
 \left(a^{}_{\rm u}\tilde{b}^{}_{\rm u}-b_{\rm u}^2\right)
&\hspace{0.2cm}c^{}_{\rm u} b^{}_{\rm u} \rho
+\displaystyle \frac{\rho -1}{\Delta^{}_1+\Delta^{}_2}
c^{}_{\rm u}b^{}_{\rm u}
\left(a^{}_{\rm u}\tilde{b}^{}_{\rm u}-b_{\rm u}^2\right)\\
 \cdots &\hspace{0.2cm}
 \begin{array}{c}
 \vspace{0.2cm}
  c_{\rm u}^2+\Delta^{}_2\rho^{2}
 -2\displaystyle \frac{\rho^{2}-\rho}{\Delta^{}_1+\Delta^{}_2}
\left(a^{}_{\rm u}\tilde{b}^{}_{\rm u}-b_{\rm u}^2\right)^2
 \end{array}
 &\hspace{0.2cm}\Delta^{}_3\rho^{2}\\
\cdots &\hspace{0.2cm}\cdots &\hspace{0.2cm}
\begin{array}{c}
 \vspace{0.2cm}
\Delta^{}_1\rho^{2}
-2\displaystyle \frac{\rho^{2}-\rho}{\Delta^{}_1+\Delta^{}_2}
\left(a^{}_{\rm u}\tilde{b}^{}_{\rm u}-b_{\rm u}^2\right)^2
\end{array}
 \end{array}
 \right) \; ,
\end{equation}
\normalsize
where the elements denoted as ``$\cdots$" can be directly read off
by considering the Hermiticity of $H^{}_{\rm u}$,
and $\rho$ describes the RGE running effects from $\Lambda^{}_{\rm FS}$
to $M^{}_Z$:
\begin{equation}\label{}
  \rho=\exp\left\{\displaystyle \frac{3}{32\pi^2}\int^{0}_{t^{}_{\rm FS}}
  y^2(t^\prime) {\rm d} t^\prime\right\}\ .
\end{equation}
Here $t^{}_{\rm FS}=\ln\left(\Lambda^{}_{\rm FS}/M^{}_Z\right)$,
and $y(t^\prime)$ is the Yukawa coupling eigenvalue of the top quark
which evolves according to
\begin{equation}\label{}
  8\pi^2 \displaystyle \frac{{\rm d} y^2}{{\rm d} t}=\left(\displaystyle
  \frac{9}{2}y^2-G^{}_{\rm u}\right)y^2\ .
\end{equation}
For illustration, $\rho \sim 0.9$ when $\Lambda^{}_{\rm FS} \sim 10^{15}$
GeV, as shown in FIG.~\ref{sigmarho}. On the other hand,
\footnotesize
\begin{equation}\label{}
H^{}_{\rm d}(M^{}_{Z}) \simeq \left(
 \begin{array}{ccc}
 \vspace{0.3cm}
 \left|c^{}_{\rm d}\right|^2 &c^{}_{\rm d}\tilde{b}^{}_{\rm d} \rho^{-1}
 -\displaystyle \frac{\rho^{-1}-1}{\Delta^{}_1+\Delta^{}_2}c^{}_{\rm d}
 \left(\Delta^{}_1\tilde{b}^{}_{\rm d}-\Delta^{}_3 b^{}_{\rm d}\right)
&c^{}_{\rm d} b^{}_{\rm d} \rho^{-1} -\displaystyle
\frac{\rho^{-1}-1}{\Delta^{}_1+\Delta^{}_2}c^{}_{\rm d}\left(\Delta^{}_2
b^{}_{\rm d} -\Delta^{}_3 \tilde{b}^{}_{\rm d}\right)\\
   \vspace{0.3cm}
 \cdots &
 \begin{array}{c}
  \left|c^{}_{\rm d}\right|^2+\left(\left|b^{}_{\rm d}\right|^2+
  \tilde{b}_{\rm d}^2\right)\rho^{-2}
  +\displaystyle \frac{\rho^{- 2}-\rho^{-1}}{\Delta^{}_1+\Delta^{}_2}\\
 \times \left[\Delta^{}_3 \left(a^{}_{\rm d}+\tilde{b}^{}_{\rm d}\right)
 \left(b^{}_{\rm d}+b_{\rm d}^*\right)
 -2\Delta^{}_1 \left(\left|b^{}_{\rm d}\right|^2+\tilde{b}_{\rm d}^2
 \right)\right]
 \end{array}
 &\begin{array}{c}
 b^{}_{\rm d}\left(a^{}_{\rm d}+\tilde{b}^{}_{\rm d}\right)\rho^{-1}
 +\displaystyle
 \frac{\rho^{- 2}-\rho^{-1}}{\Delta^{}_1+\Delta^{}_2}\\
 \times \left(a_{\rm d}^2+2\left|b^{}_{\rm d}\right|^2+
 \tilde{b}_{\rm d}^2\right)\Delta^{}_3
  \end{array}\\
\cdots &\cdots &
\begin{array}{c}
 \left(a_{\rm d}^2+\left|b^{}_{\rm d}\right|^2\right)\rho^{-2}
 +\displaystyle
 \frac{\rho^{-2}-\rho^{-1}}{\Delta^{}_1+\Delta^{}_2}\\
 \times \left[\Delta^{}_3 \left(a^{}_{\rm d}+\tilde{b}^{}_{\rm d}\right)
 \left(b^{}_{\rm d} +b_{\rm d}^*\right) -2\Delta^{}_2
 \left(a_{\rm d}^2+\left|b^{}_{\rm d}\right|^2\right)\right]
\end{array}
 \end{array}
 \right)
\end{equation}
\normalsize
The RGE-corrected quark mass matrices
can then be extracted from Eqs. (59) and (62):
\footnotesize
\begin{eqnarray}\label{mumz}
M^{}_{\rm u}(M^{}_Z) & \simeq & \sigma^{}_{\rm u}v\left(
 \begin{array}{ccc}
 \vspace{0.3cm}
 0 &\hspace{0.3cm}c^{}_{\rm u}&\hspace{0.3cm}0\\
   \vspace{0.3cm}
 \cdots &\hspace{0.3cm}
 \begin{array}{c}
 \tilde{b}^{}_{\rm u}\rho -\displaystyle \frac{\rho-1}{\Delta^{}_1
 +\Delta^{}_2} a^{}_{\rm u}\left(a^{}_{\rm u}\tilde{b}^{}_{\rm u}
 -b_{\rm u}^2\right)
 \end{array}
 &\hspace{0.3cm}\begin{array}{c}
  b^{}_{\rm u}\rho+\displaystyle \frac{\rho-1}{\Delta^{}_1
  +\Delta^{}_2} b^{}_{\rm u}\left(a^{}_{\rm u}\tilde{b}^{}_{\rm u}
  -b_{\rm u}^2\right)
  \end{array}\\
\cdots &\hspace{0.3cm}\cdots &\hspace{0.3cm}
\begin{array}{c}
 a^{}_{\rm u}\rho-\displaystyle \frac{\rho-1}{\Delta^{}_1
 +\Delta^{}_2}\tilde{b}^{}_{\rm u}\left(a^{}_{\rm u}
 \tilde{b}^{}_{\rm u}-b_{\rm u}^2\right)
\end{array}
 \end{array}
 \right) \; , \nonumber \\
M^{}_{\rm d}(M^{}_Z) & \simeq & \sigma^{}_{\rm d} v\left(
 \begin{array}{ccc}
 \vspace{0.3cm}
 0 &\hspace{0.3cm}c_{\rm d}&\hspace{0.3cm}0\\
   \vspace{0.3cm}
 c_{\rm d}^* &\hspace{0.3cm}
  \tilde{b}^{}_{\rm d} \rho^{-1} +\displaystyle
  \frac{\rho^{-1}-1}{\Delta^{}_1+\Delta^{}_2}\left(\Delta^{}_3
  b_{\rm d}^*-\Delta^{}_1 \tilde{b}^{}_{\rm d}\right)
&\hspace{0.3cm} b^{}_{\rm d}\rho^{-1}+\displaystyle
\frac{\rho^{-1}-1}{\Delta^{}_1+\Delta^{}_2} \left(\Delta^{}_3
a^{}_{\rm d}-\Delta^{}_1 b^{}_{\rm d}\right)
\\ 0 &\hspace{0.3cm}
b_{\rm d}^*\rho^{-1}+\displaystyle \frac{\rho^{-1}-1}{\Delta^{}_1
+\Delta^{}_2}\left(\Delta^{}_3 \tilde{b}^{}_{\rm d}-
\Delta^{}_2 b_{\rm d}^*\right) &\hspace{0.3cm}
 a^{}_{\rm d}\rho^{{-1} }+\displaystyle \frac{\rho^{-1}-1}
 {\Delta^{}_1+\Delta^{}_2}
\left(\Delta^{}_3 b^{}_{\rm d}-\Delta^{}_2 a^{}_{\rm d}\right)
 \end{array}
 \right) \; , \hspace{0.6cm}
\end{eqnarray}
\normalsize
where
\begin{equation}\label{}
  \sigma^{}_{\rm q}=\exp\left\{\displaystyle \frac{1}
  {16\pi^2}\int^{0}_{t^{}_{\rm FS}}\left[3y^2(t^\prime)-G^{}_{\rm q}
  (t^\prime)\right] {\rm d} t^\prime\right\}\
\end{equation}
is the overall rescaling factor of quark mass matrices brought back
from the $G_{\rm q}(t)$ and $T(t)$ terms of Eq.~(\ref{rge}) that
were tentatively dropped in Eq.~(\ref{rge2}).
Apparently, $M^{}_{\rm d}(M_Z)$ is not Hermitian any more,
because the RGE of $Y^{}_{\rm d}(t)$ does not respect
Hermiticity. To illustrate, the numerical changes of $\sigma^{}_{\rm u}$
and $\sigma^{}_{\rm d}$ with the scale $\Lambda_{\rm FS}$ are
shown in FIG.~\ref{sigmarho} in the framework of the SM. Of course,
the above analytical results can exactly reproduce those obtained
in Ref. \cite{lindner} for the Fritzsch ansatz of quark mass matrices
when $\tilde{b}^{}_{\rm u}$ and $\tilde{b}^{}_{\rm d}$
are switched off.

Note that the geometrical relation in Eq.~(\ref{abtb}) can be
reexpressed as $(a^{}_{\rm q}\tilde{b}^{}_{\rm q}-
b_{\rm q}^2)/(\Delta_1^{}+\Delta_{2}^{}) \sim m^{}_2/m^{}_3$.
Hence in each entry of the (2,3)
sector of $M^{}_{\rm u}(M^{}_Z)$ the second term
is suppressed by a factor proportional to
$(1-\rho)\ m^{}_c/m^{}_t \lesssim 10^{-3}$
as compared with the first term. As for $M^{}_{\rm d}(M^{}_Z)$,
let us take its (2,3) entry as an example to look at the corresponding
RGE correction. Because of the parallelism between $(a^{}_{\rm
u}, b^{}_{\rm u}, \tilde{b}^{}_{\rm u})$ and $(a^{}_{\rm d},
{\rm Re}(b^{}_{\rm d}), \tilde{b}^{}_{\rm d})$, we find
\begin{eqnarray}\label{}
  \Delta^{}_3 a^{}_{\rm d}-\Delta^{}_1 {\rm Re}(b^{}_{\rm d})
  & = & \Delta^{}_1\rm Re(b^{}_{\rm d})
  \left[\displaystyle \frac{\Delta^{}_3}{\Delta^{}_1}
  \frac{a^{}_{\rm d}}{{\rm Re}(b^{}_{\rm d})}-1\right] \nonumber \\
  & \simeq & \Delta^{}_1 {\rm Re}(b^{}_{\rm d}) \left[\displaystyle
  \frac{\Delta^{}_3}{\Delta^{}_1}\frac{a^{}_{\rm u}}
  {b^{}_{\rm u}}-1\right]
  ={\rm Re}(b^{}_{\rm d})\left(a^{}_{\rm u}\tilde{b}^{}_{\rm u}-b_{\rm u}^2
  \right) \ .
\end{eqnarray}
So the real part of the second term of $M^{}_{\rm d 23}$ at $M^{}_Z$
is suppressed by a factor proportional to
$(1-\rho)\ m^{}_c/m^{}_t \lesssim 10^{-3}$ as compared with the real
part of its first term. In other words,
the real part of $M^{}_{\rm d 23}$ is approximately equal
to $\rho^{-1} {\rm Re}(b^{}_{\rm d})$ at $M^{}_Z$.
According to our phase assignment in Eq.~(\ref{repara}),
$\phi^{}_2 = \arg(b^{}_{\rm u}) - \arg(b^{}_{\rm d})
= -\arg(b^{}_{\rm d})$ holds.
Hence the phase of $b^{}_{\rm d}$ is equal to $-\phi^{}_2$ and
must be close to $0$ or $-2\pi$. In the $r \sim 1$ region where
$\Delta^{}_2$ is much smaller than $\Delta^{}_1$,
it is easy to see that the imaginary part of the (2,3)
entry of $M^{}_{\rm d}(M^{}_Z)$ is about ${\rm Im}(b^{}_{\rm d})$.
That means $\arg(b^{}_{\rm d}) \simeq
{\rm Im}(b^{}_{\rm d})/{\rm Re}(b^{}_{\rm d})$ is rescaled
by $\rho$ due to the RGE effects, or equivalently
\begin{equation}\label{phi2}
  \phi^{}_{2}(\Lambda^{}_{\rm FS})\simeq \rho^{-1}
  \phi^{}_{2}(M^{}_{Z})\ .
\end{equation}
In a word, the four texture zeros of quark mass matrices
are essentially stable against the evolution of energy scales. To be
more specific, $M^{}_{\rm u}$ and $M^{}_{\rm d}$
develop the overall factors $\sigma^{}_{\rm u}$ and $\sigma^{}_{\rm d}$
during their running from $\Lambda^{}_{\rm FS}$ down to $M^{}_Z$,
respectively; and their finite entries $(a^{}_{\rm
u}, b^{}_{\rm u}, \tilde b^{}_{\rm u})$ and $(a^{}_{\rm d},
{\rm Re}(b^{}_{\rm d}), \tilde{b}^{}_{\rm d})$ are rescaled by $\rho$
and $\rho^{-1}$, respectively.

To illustrate the RGE-induced corrections, let us give a numerical
example to compare between Eq. (\ref{repara})
at $\Lambda^{}_{\rm FS}$ and Eq. (\ref{mumz}) at $M^{}_Z$.
We first figure out the values of quark masses and flavor mixing
parameters at $\Lambda^{}_{\rm {FS}} \sim 10^{11}$ GeV by solving
the one-loop RGEs numerically:
\begin{equation}\label{}
 \begin{array}{lll}
  m^{}_{u}=0.69\ {\rm MeV} \; , &\hspace{1cm}m^{}_c=320\ {\rm MeV} \; ,
  &\hspace{1cm}m^{}_t=95.6\  {\rm GeV} \; ;\\
  m^{}_d=1.4\ {\rm MeV} \; , &\hspace{1cm}m^{}_s=29.1\ {\rm MeV} \; ,
  &\hspace{1cm}m^{}_b=1.3\ {\rm GeV} \; ;\\
  |V^{}_{us}|=0.225\ , &\hspace{1cm} |V^{}_{cb}|=0.0458\ ,
  &\hspace{1cm} |V^{}_{ub}|=0.00387 \ ,
  \end{array}
\end{equation}
and the value of $\sin 2\beta$ is almost unchanged from $M^{}_Z$ to
$\Lambda^{}_{\rm FS}$ (or vice versa) within the accuracy that we
need. The choice of this specific scale is for two simple reasons:
on the one hand, it is expected to be around the canonical seesaw
\cite{SS} and leptogenesis \cite
{FY} scales; on the other hand, it
is close to the energy scale relevant for the possible vacuum
stability issue of the SM \cite{stability}. Therefore,
\begin{eqnarray}\label{}
Y^{}_{\rm u}(\Lambda^{}_{\rm FS}) & \simeq & 10^{-1} \left(
 \begin{array}{ccc}
 0&\hspace{0.1cm} 9\times 10^{-4} &\hspace{0.1cm} 0\\
\cdots &\hspace{0.1cm} 0.6 & \hspace{0.1cm}1.7\\
\cdots&\hspace{0.1cm} \cdots &\hspace{0.1cm}4.9
 \end{array}
 \right) \; , \nonumber \\
Y^{}_{\rm d}(\Lambda^{}_{\rm FS}) & \simeq & 10^{-3}\left(
 \begin{array}{ccc}
 0&\hspace{0.1cm} 0.04\ e^{-1.67 \rm i} &\hspace{0.1cm} 0\\
\cdots &\hspace{0.1cm} 1.0 & \hspace{0.1cm}2.4\ e^{0.14 \rm i}\\
\cdots&\hspace{0.1cm}\cdots &\hspace{0.1cm} 6.7
 \end{array}
 \right) \; .
\end{eqnarray}
In comparison, the corresponding quark mass matrices at the
electroweak scale are
\small
\begin{eqnarray}\label{}
M^{}_{\rm u}(M^{}_Z) & \simeq & 10^{-1} \sigma^{}_{\rm u}v\left(
 \begin{array}{ccc}
 \vspace{0.2cm}
 0 &\hspace{0.3cm}9\times 10^{-4}&\hspace{0.3cm}0\\
   \vspace{0.2cm}
 \cdots &\hspace{0.3cm} 0.6\rho - 8\cdot10^{-3} \left(\rho-1\right)
 &\hspace{0.3cm} 1.7\rho + 3\times 10^{-3}\left(\rho-1\right) \\
\cdots &\hspace{0.3cm}\cdots &\hspace{0.3cm} 4.9\rho
-10^{-3} \left(\rho-1\right)
 \end{array}
 \right) \; , \nonumber \\
M^{}_{\rm d}(M^{}_Z) & \simeq & 10^{-3}\sigma^{}_{\rm d} v\left(
 \begin{array}{ccc}
 \vspace{0.2cm}
 0 &0.04\ e^{-1.67 \rm i}&0\\
   \vspace{0.2cm}
 \cdots &1.0\rho^{-1} -\left(0.2+0.1 \rm i\right)\varepsilon &
 2.4\rho^{-1}+0.3 {\rm i} -0.04\varepsilon\\
\cdots & \left(2.4-0.3 {\rm i}\right)\rho^{-1}
+ \left(0.05+0.04 {\rm i}\right)\varepsilon &
\hspace{0.2cm}6.7\rho^{-1}+\left(0.01+0.1 {\rm i}\right)\varepsilon
 \end{array}
 \right) \; , \hspace{0.5cm}
\end{eqnarray}
\normalsize where $\varepsilon = \rho^{-1}-1$ is a small value of
${\cal O}(0.1)$ or much smaller. This numerical exercise confirms
our qualitative analysis made above. In particular, the imaginary
part of the (2,3) entry of $M^{}_{\rm d}(M^{}_Z)$ is really
independent of $\rho$, and its real part is proportional to
$\rho^{-1}$.

Now let us turn to the running behaviors of quark masses and flavor
mixing parameters. Since $c^{}_{\rm u}$ is negligibly small in
magnitude as compared with $a^{}_{\rm u}$, $b^{}_{\rm u}$ and
$\tilde{b}^{}_{\rm u}$, the invariants of the (2,3) submatrix of
$M^{}_{\rm u}(\Lambda^{}_{\rm FS})$ and $M^{}_{\rm u}(M^{}_{Z})$
lead us to
\begin{eqnarray}\label{}
&&  m^{}_{c}(\Lambda^{}_{\rm FS})+ m^{}_{t}(\Lambda^{}_{\rm FS})
\simeq   v\left(a^{}_{\rm u} +\tilde{b}^{}_{\rm u}\right)\ , \nonumber \\
&&  m^{}_{c}(\Lambda^{}_{\rm FS})\  m^{}_{t}(\Lambda^{}_{\rm FS})
\simeq   v^2\left(a^{}_{\rm u}\tilde{b}^{}_{\rm u}-b_{\rm
u}^2\right) \ ; \nonumber \\
&&  m^{}_{c}(M^{}_{Z})+ m^{}_{t}(M^{}_{Z}) \simeq \sigma^{}_{\rm u}
v \left(a^{}_{\rm u} +\tilde{b}^{}_{\rm u}\right)\rho\ , \nonumber \\
&&  m^{}_{c}(M^{}_{Z})\  m^{}_{t}(M^{}_{Z}) \simeq \sigma_{\rm u}^2
v^2 \left(a_{\rm u} \tilde{b}^{}_{\rm u}-b_{\rm u}^2 \right)\rho\ .
\end{eqnarray}
These relations indicate that $m^{}_c$ and $m^{}_t$ change with the
energy scale in the following way:
\begin{equation}\label{mtc}
 m^{}_{t}(\Lambda^{}_{\rm FS}) \simeq \sigma_{\rm u}^{-1}\rho^{-1}
 m^{}_{t}(M^{}_{Z})\ ,
 \hspace{1cm}
  m^{}_{c}(\Lambda^{}_{\rm FS}) \simeq \sigma_{\rm u}^{-1} m^{}_{c}(M^{}_{Z})\ .
\end{equation}
When $c^{}_{\rm u}$ is concerned, a similar trick yields
$m^{}_{u}(\Lambda^{}_{\rm FS}) \simeq \sigma_{\rm u}^{-1}
m^{}_{u}(M^{}_{Z})$. It is easy to verify that the similar relations
hold in the down sector:
\begin{equation}\label{mbsd}
 m^{}_{b}(\Lambda^{}_{\rm FS}) \simeq
 \sigma_{\rm d}^{-1}\rho \ m^{}_{b}(M^{}_{Z})\ ,
 \hspace{1cm} m^{}_{s}(\Lambda^{}_{\rm FS}) \simeq
 \sigma_{\rm d}^{-1} m^{}_{s}(M^{}_{Z})\ ,
\end{equation}
and $m^{}_{d}(\Lambda^{}_{\rm FS}) \simeq\sigma_{\rm d}^{-1}
m^{}_{d}(M^{}_{Z})$. These results clearly show that the mass ratios
$m^{}_u/m^{}_c$ and $m^{}_d/m^{}_s$ are essentially free from the
RGE corrections.

To see how the flavor mixing parameters evolve from $\Lambda^{}_{\rm
FS}$ down to $M^{}_Z$, we take a new look at Eq.~(\ref{ckmelement}).
Above all, the dimensionless parameters $r^{}_{\rm u}$ and
$r^{}_{\rm d}$ are independent of the energy scale to a good degree
of accuracy. The reason is simply that $m^{}_t$ (or $m^{}_b$) and
$A^{}_{\rm u}$ (or $A^{}_{\rm d}$) have nearly the same running
behaviors, as one can see from Eqs. (63), (71) and (72). It is also
straightforward to conclude that $|V^{}_{us}|$ is stable against the
evolution of energy scales. In view of $r^{}_{\rm u}\simeq r^{}_{\rm
d}$ and $\phi^{}_2\simeq 0$, we arrive at the approximation
\begin{equation}\label{}
  |V^{}_{cb}|\simeq \displaystyle \sqrt{\left(1-r^{}_{\rm u}\right)
  r^{}_{\rm d}}\left| \displaystyle \frac{1}{2}\
  \displaystyle \frac{\eta^{}_{\rm d}}{r^{}_{\rm d}}\ \displaystyle
  \frac{m^{}_s}{m^{}_b}- {\rm i} \sin{\phi^{}_2} \right|\ .
\end{equation}
Given Eqs.~(\ref{phi2}) and (\ref{mbsd}), the running behavior of
$|V^{}_{cb}|$ turns out to be
\begin{equation}\label{}
  \left|V^{}_{cb}(\Lambda^{}_{\rm FS})\right|
  \simeq \rho^{-1}\left|V^{}_{cb}(M^{}_{Z})\right| \; .
\end{equation}
With the help of this result and Eq.~(\ref{ckmelement}), we
immediately obtain
\begin{equation}\label{}
  \left|V^{}_{ub}(\Lambda^{}_{\rm FS})\right| \simeq \rho^{-1}
  \left|V^{}_{ub}(M^{}_{Z})\right| \; .
\end{equation}
In addition, Eq.~(\ref{beta}) tells us that $\beta$ is nearly
scale-independent. It is easy to check that $\alpha$ and $\gamma$,
the other two inner angles of the CKM unitarity triangle, are also
free from the RGE corrections at the one-loop level \cite{Luo}.

The above results can simply be translated into the ones for three
flavor mixing angles and one CP-violating phase in the standard
parametrization of the CKM matrix:
\begin{equation}\label{angles}
\begin{array}{ll}
  \theta^{}_{12}(\Lambda^{}_{\rm FS})\simeq \theta^{}_{12}(M^{}_{Z})\ ,
  &\hspace{1cm}
  \theta^{}_{23}(\Lambda^{}_{\rm FS})\simeq \rho^{-1}\theta^{}_{23}(M^{}_{Z})\
  ,\\
  \theta^{}_{13}(\Lambda^{}_{\rm FS})\simeq \rho^{-1}\theta^{}_{13}(M^{}_{Z})\
  ,&\hspace{1cm}
   \delta(\Lambda^{}_{\rm FS})\simeq \delta(M^{}_{Z})\ .
\end{array}
\end{equation}
Of course, $\alpha$, $\beta$ and $\gamma$ are all the functions of
$\delta$ in this parametrization. As for the Jarlskog invariant
${\cal J} = \cos\theta^{}_{12} \sin\theta^{}_{12}
\cos^2\theta^{}_{13} \sin\theta^{}_{13} \cos\theta^{}_{23}
\sin\theta^{}_{23}\sin\delta$ \cite{J}, it is easy to arrive at
${\cal J}(\Lambda^{}_{\rm FS}) \simeq \rho^{-2} {\cal J}(M^{}_{Z})$
in the same approximation. Such a rephasing-invariant measure of
weak CP violation is actually tiny, only about $3\times 10^{-5}$ at
$M^{}_Z$.

Finally, let us briefly comment on a possible implication of the
loss of Hermiticity of $M^{}_{\rm d}$ running from $\Lambda^{}_{\rm
FS}$ down to $M^{}_Z$. We conjecture that it might have something to
do with the strong CP problem \cite{scp}, which is put forward due
to the {\it unnatural} smallness of the parameter $\overline{
\theta} = \theta^{}_{\rm QCD}+\theta^{}_{\rm QFD}$. Here
$\theta^{}_{\rm QCD}$ is the coefficient of the CP-violating term in
the QCD Lagrangian \cite{qcd},
\begin{equation}\label{}
 \mathcal L^{}_{\theta}=\theta^{}_{\rm QCD} \displaystyle
 \frac{g_{3}^2}{32 \pi^2}G^{}_{\mu\nu} \tilde{G}^{\mu\nu}\ ;
\end{equation}
and $\theta^{}_{\rm QFD}$ comes from the quark flavor sector,
\begin{equation}\label{}
 \theta^{}_{\rm QFD}=\arg\left[{\rm Det}\left(M^{}_{\rm u}
 M^{}_{\rm d}\right)\right] \ .
\end{equation}
The experimental upper bound of $\overline \theta$ is at the
$10^{-11}$ level \cite{theta}, in sharp contrast with a {\it
natural} value of $\mathcal O(1)$ from a theoretical point of view.
The demand for explaining why $\overline \theta$ is so tiny poses
the strong CP problem. An attractive solution for this problem is
the Peccei-Quinn mechanism \cite{pq} in which an anomalous $U(1)$
symmetry is introduced to ensure a complete cancellation between
$\theta_{\rm QCD}$ and $\theta_{\rm QFD}$. Another competitive
strategy is to remove $\theta_{\rm QCD}$ by imposing a spontaneously
broken P or CP symmetry (e.g., in the LR symmetric model), and to
keep the second term vanishing in the meantime
\cite{sponcp,sponcp2}.

Being Hermitian, the four-zero texture of quark mass matrices
automatically satisfies the requirement $\arg\left[{\rm
Det}\left(M^{}_{\rm u} M^{}_{\rm d}\right)\right] =0$ at a superhigh
energy scale $\Lambda^{}_{\rm FS}$. Nevertheless, the RGE effects
can render $M^{}_{\rm d}(M^{}_Z)$ non-Hermitian as shown in
Eq.~(\ref{mumz}). Since the strong CP term begins to take effect at
the scale of about 260 MeV where the QCD vacuum transforms, nonzero
$\arg\left[{\rm Det}\left(M^{}_{\rm d}\right)\right]$ at or below
$M^{}_Z$ will contribute to $\overline\theta$ in spite of
$\arg\left[{\rm Det}\left(M^{}_{\rm d}\right)\right] =0$ at
$\Lambda^{}_{\rm FS}$. Given the explicit form of $M^{}_{\rm
u}(M^{}_Z)$ and $M^{}_{\rm d}(M^{}_Z)$ in Eq. (63), one may
calculate its contribution to $\overline\theta$ as follows:
\begin{equation}\label{arg}
  \theta^{}_{\rm QFD}=\arg\left[{\rm Det}M^{}_{\rm d}(M^{}_Z)\right]
  \simeq \arctan \left[\frac{\rho^{-1}-1}{\Delta^{}_1+\Delta^{}_2}
\frac{\Delta^{}_3 \ {\rm Im}(b^{}_{\rm d})}{a^{}_{\rm d}\
\rho^{-1}}\right] \sim \left(1-r\right)^2 \left(1-\rho\right)
\sin{\phi^{}_2} \ .
\end{equation}
Although $\phi^{}_2$ is very small, it cannot be exactly zero as
shown in our numerical analysis. Given $r=0.9$ and $\Lambda^{}_{\rm
FS}=1$ TeV, for instance, Eq. (79) leads us to a value of ${\cal
O}(10^{-5})$, much larger than the upper bound of $\overline\theta$.
One way out of this problem is to fine-tune the value of $r$. But
the possibility of $r \simeq 1$ has phenomenologically been ruled
out, as discussed at the beginning of section \Rmnum 3. If the
parallelism between the forms of $M^{}_{\rm u}$ and $M^{}_{\rm d}$
is given up, the situation will change. For example, in a flavor
basis with $M^{}_{\rm u}$ being diagonal, the value of
$\theta^{}_{\rm QFD}$ was estimated to be of ${\cal O}(10^{-16})$ in
Ref.~\cite{qfd}. In short, it seems difficult to directly employ the
four-zero texture of quark mass matrices to solve the strong CP
problem in the scenario of spontaneous CP violation. But a more
detailed study of this issue is needed before a firm conclusion can
be achieved.

\section{Summary}

We have carried out a new study of the four-zero texture of
Hermitian quark mass matrices and improved the previous works in
several aspects. In our numerical analysis what really matters is
that we have found a new part of the parameter space, corresponding
to $A \sim |B|\sim \tilde B$ (or $r\sim0.5$), and confirmed the
known part corresponding to $A > |B|> \tilde B$ (or $r\sim 1$). In
particular, the exact equality between $A$ and $\tilde B$ is
allowed, and this opens an interesting window for model building.
We want to emphasize that the newly found parameter space is
phenomenologically different from the already known: since the
former allows the (near) equality of mass matrix entries ---
a characteristic of non-Abelian flavor symmetries which are very
popular in the lepton sector \cite{f-symmetry}, it provides a
possibility of unifying the description of quarks and leptons with
the same flavor symmetries and this will be discussed elsewhere.

We have identified two special four-zero patterns of quark mass
matrices and constructed two toy flavor-symmetry models to realize
them. One of the patterns possesses a mild hierarchy $A\sim \epsilon
|B|\sim \epsilon^2 \tilde B$ with $\epsilon$ being about 0.3, and it
can be obtained with the help of the FN mechanism. The other pattern
assumes $A=\tilde B$, which can be realized by means of the $S(3)$
flavor symmetry. Both of them show a similarity between the (2,3)
sectors of $M^{}_{\rm u}$ and $M^{}_{\rm d}$, indicating that the
latter could have the same origin. We have done two model-building
exercises in the SUSY LR framework with an explicit parity symmetry,
which ensures the Hermiticity of quark mass matrices at the flavor
symmetry scale $\Lambda^{}_{\rm FS}$.

We have also studied the RGE effects on the four-zero texture of
quark mass matrices in an analytical way, from $\Lambda^{}_{\rm FS}$
down to the electroweak scale $M^{}_Z$. Our results show that the
texture zeros of $M^{}_{\rm u}$ and $M^{}_{\rm d}$ are essentially
stable against the evolution of energy scales, but their finite
entries are rescaled due to the RGE-induced corrections. An
interesting consequence of the RGE running is the loss of the
Hermiticity of $M^{}_{\rm d}$ at $M^{}_Z$ in the SM. As a byproduct,
the possibility of applying the four-zero texture of quark mass
matrices to resolving the strong CP problem has been discussed in a
very brief way.

Although the predictive power of texture zeros has recently been
questioned in the lepton sector \cite{Grimus}, they
remain useful in the quark sector to understand the
correlation between the hierarchy of quark masses and that of
flavor mixing angles. We remark that possible flavor symmetries
are behind possible texture zeros, and they are phenomenologically
important to probe the underlying flavor structure before a
complete flavor theory is developed.

\vspace{0.5cm}
{\it Note added.} While our paper was being finished, we noticed
a new preprint \cite{LG} in which a systematic survey of possible
texture zeros of quark mass matrices was done but the four-zero
texture of Hermitian quark mass matrices with
the up-down parallelism was not explicitly discussed.

\begin{acknowledgments}

This work was supported in part by the National Natural Science
Foundation of China under Nos. 11375207 and 11135009.

\end{acknowledgments}

\end{document}